# A Note on the Periodicity and the Output Rate of Bit Search Type Generators

Yücel Altuğ,  N. Polat Ayerden,  M. Kıvanç Mıhçak *Member* and Emin Anarım

### Abstract

We investigate the bit-search type irregular decimation algorithms that are used within LFSR-based stream ciphers. In particular, we concentrate on BSG and ABSG, and consider two different setups for the analysis. In the first case, the input is assumed to be a $m$-sequence; we show that all possible output sequences can be classified into two sets, each of which is characterized by the equivalence of their elements up to shifts. Furthermore, we prove that the cardinality of each of these sets is equal to the period of one of its elements and subsequently derive the first known bounds on the expected output period (assuming that no subperiods exist). In the second setup, we work in a probabilistic framework and assume that the input sequence is evenly distributed (i.e., independent identically distributed Bernoulli process with probability $1/2$). Under these assumptions, we derive closed-form expressions for the distribution of the output length and the output rate, which is shown to be asymptotically Gaussian-distributed and concentrated around the mean with exponential tightness.

## I. Introduction

Within symmetric key encryption, there are two main classes of schemes: Block ciphers and stream ciphers. As far as stream ciphers are concerned, the usage of linear feedback shift registers (LFSRs) as the main building block is quite common in practice because of the implementation efficiency, speed and good statistical properties of the output. It is well-known that the cryptanalysis of LFSRs is of polynomial complexity due to their linearity properties [1]. Therefore, it is essential to bring additional non-linearities to the LFSR outputs in order to enhance the security of the resulting system. One such approach includes applying irregular decimation techniques to the LFSR output [2], [3], [4], [5]. Such techniques may render several conventional attacks useless, such as algebraic attacks, which are known to be one of the most effective attack algorithms designed against LFSR-based stream ciphers. We use the term "decimation-type algorithms" to denote algorithms that use irregular decimation techniques.

Shrinking [4] and self-shrinking generators (SSG) [5] are two important examples of this class; in the literature, they are known to be "pioneering" algorithms that employ the idea of "decimation". SSG is very simple and efficient in terms of hardware implementation; furthermore, its security against known attacks makes it one of the most popular state-of-the-art decimation-type algorithms [6], [7], [8]. In SSG, the main idea is to split the input bitstream into blocks of length-2 and produce the output bit stream as a function of the first bit of the input blocks.

Bit-search generator (BSG) [2] and its subsequent variant ABSG [3] are relatively newer techniques, which also qualify as decimation-type algorithms. In contrast with SSG, the approach in bit-search type methods is to "look for" particular patterns of variable lengths in order to produce an output bit; the type of the particular mapping that operates on the input which produces the output stream determines the difference between BSG and ABSG. A detailed comparison between ABSG and BSG is given in [3]. ABSG and BSG have the same asymptotic output rate, which can be shown to be better than SSG. Moreover, it has been shown in [9] that, against most known attacks, ABSG is the "best" choice among a wide class of decimation-type algorithms in the sense of known attack complexities. Therefore, we believe that ABSG is worth investigating further, which basically forms the essence of this paper.

**Focus of the paper:** As far as stream ciphers are concerned, characterization of rate and periodicity is of fundamental significance. Qualitatively, the "rate" (i.e., output rate) of a decimation-type algorithm is defined as the reciprocal of the number of input bits to produce one bit of output, on average. Therefore, the rate of a decimation-type algorithm directly determines the efficiency of the resulting stream cipher. Next, it is well-known that (e.g., see [10]) the "period" (i.e., output period) of a LFSR-based stream cipher is required to be large as a necessary condition for the security of the system [1]. In this paper, we focus on the analytical quantification of rate and periodicity of bit-search type generators, namely BSG [2] and ABSG [3].

The authors are with the Electrical and Electronic Engineering Department of Boğaziçi University, Istanbul, 34342, Turkey (e-mail: yucel.altug@boun.edu.tr, nafizpol@boun.edu.tr,  kivanc.mihcak@boun.edu.tr,  anarim@boun.edu.tr )

N. P. Ayerden and E. Anarım are partially supported by the State Planning Organization of Turkey under grant no. 2003K120250; M. K. Mıhçak is partially supported by TÜBİTAK Career Award no. 106E117.

[1]Throughout this work, when we refer to the concept of "period", we essentially imply the "least period".

**Prior results:** In [2], the authors have intuitively argued that the asymptotic rate of the BSG algorithm is $1/3$. Moreover, based on experimental evidence, they have conjectured that, for $m$-sequences (i.e., maximal-length LFSR outputs which have been generated using a primitive feedback polynomial [1]), BSG algorithm produces exactly two types of output sequences in terms of their periods, namely with approximate periods of $T/3$ and $2T/3$. Further, based on this observation, they have argued that the average resulting period is always $T/2$. In [3], based on the assumption that the input sequence is a realization i.i.d. (independent identically distributed) Bernoulli process with probability $1/2$ (i.e., the input sequence is "evenly distributed"), the authors have also mentioned that the expected output rate is $1/3$. Arguments about the output period of ABSG, which are similar to the ones presented in [2] are also mentioned in [3].

**Our contribution:** We derive analytical results on the rate and periodicity properties of BSG and ABSG algorithms in deterministic and probabilistic setups:

- Deterministic Setup: In this case, we assume that the input sequence is a $m$-sequence.
  - We prove that both BSG and ABSG produce *exactly* two types of output sequences with respect to their periods. In particular, the set of output sequences is given as the union of two disjoint sets; within each one of these sets, the elements are equivalent to each other up to shifts.
  - We show that the cardinality of each one of the aforementioned two sets is equal to the period of any element included in that set. Using this result, we derive the first known bounds on the expected period of the output sequence of BSG and ABSG algorithms under the assumption that no subperiods exist.

- Probabilistic Setup: In this framework, we assume that the input sequence is a realization of an i.i.d. Bernoulli process with probability $1/2$.
  - Given the length of the input sequence (say $N$), we derive a closed form expression for the distribution of the length of the output sequence produced by BSG and ABSG algorithms. Using this result, we analytically derive output rates of both algorithms.
  - Moreover, we prove that, the aforementioned distribution converges to a Gaussian distribution with the mean and variance of $N/3$ and $2N/27$, respectively, which, in return, implies that the concentration around the mean is exponentially tight. As a result we show that output rate is exponentially concentrated around $1/3$.

The organization of the paper is as follows. In Section II, the notation used in the paper and definitions of BSG and ABSG algorithms are given. In Section III, we derive fundamental properties of BSG and ABSG algorithms related to the periodicity under the assumption that the input is a $m$-sequence. In Section IV, we investigate the probabilistic behavior of BSG and ABSG assuming that the input is evenly distributed. The paper concludes with discussions given in Section V.

## II. Notation and Background

In this section we introduce the notation we follow throughout the paper and give basic definitions about BSG and ABSG algorithms.

### A. Notation

Boldface letters denote vectors; regular letters with subscripts denote individual elements of vectors. Furthermore, capital letters represent random variables and lowercase letters denote individual realizations of the corresponding random variable. For instance, let $\mathbf{A} \in \mathbb{R}^N$ denote a length-$N$ random vector. In that case, $A_i$ (which is a random variable) represents the $i^{\text{th}}$ entry of $\mathbf{A}$; $\mathbf{a} \in \mathbb{R}^N$ is a particular realization of $\mathbf{A}$ and similarly $a_i$ represents the $i^{\text{th}}$ entry of $\mathbf{a}$. Also, the sequence of $(a_1, a_2, \ldots, a_N)$ is compactly represented by $\mathbf{a}_1^N$. We use the notation of $(\mathbf{a}, i)$ to denote the $i$-shifted version of $\mathbf{a}$; i.e., defining $\tilde{\mathbf{a}} \triangleq (\mathbf{a}, i)$, we have $\tilde{a}_n = a_{n+i}$ for all $n$, $i \geq 0$. Furthermore, given $\mathbf{a}_1^N$, such that $a_i \in \{c_1, c_2, \ldots c_k\}$, $1 \leq i \leq N$, we define $\mathcal{W}_{c_i}\left(\mathbf{a}_1^N\right) \triangleq \sum_{j=1}^{N} \mathbf{1}_{a_j = c_i}$, where $\mathbf{1}$ denotes the standard indicator function.

### B. Description of BSG and ABSG

Both BSG and ABSG are algorithms, which are based on taking output of a pseudo-random number generator (PRNG) (e.g., LFSR) as their input, and constructing the output by irregularly decimating the input sequence. In the discussions below, let $\mathbf{x} = (x_1, x_2, \ldots)$ denote the input sequence to these algorithms.

Such an arbitrary input bit stream $\mathbf{x}$ can be partitioned into non-overlapping blocks of the form $\overline{b}, b^i, \overline{b}$, where $i \geq 0$, $b \in \{0, 1\}$, and $\overline{b}$ denotes the complement of the bit $b$. This partitioning is the common first step in both BSG and ABSG. The difference between them arises from the output bit generation mechanism once the partitioning is done. In case of BSG, an output bit is produced via $XOR$'ing the first two bits of the corresponding input block, which is of the form $\overline{b}, b^i, \overline{b}$, where $i \geq 0$. Clearly, this implies that if $i = 0$ (resp. $i > 0$), then the corresponding output bit is $0$ (resp. 1). In case of ABSG, the output bit is the second bit of the corresponding block in the input sequence; in other words, given an input block $\overline{b}, b^i, \overline{b}$, where $i \geq 0$, the output bit is $\overline{b}$ (resp. $b$) if $i = 0$ (resp. $i > 0$).

**Toy Example:** Suppose we are given the input bit stream

$$\mathbf{x} = (1, 0, 1, 0, 1, 1, 0, 0, 1, 0, 1, 1, 1, 0, 0, 0, 0, 1, 0, 0, 1, 1, 0, 1, 0, 1, 0, 0, 1, 0, 1, 1, 0).$$

- **Partitioning:** After the partitioning is done using the aforementioned rule, we have the following blocks:

$$\{(1, 0, 1), (0, 1, 1, 0), (0, 1, 0), (1, 1), (1, 0, 0, 0, 0, 1), (0, 0), (1, 1), (0, 1, 0), (0, 1, 0), (1, 0, 0, 1), (0, 1, 1, 0)\}$$

- **Output Bit Generation:**
  - <u>BSG:</u> Output bit sequence is $(1, 1, 1, 0, 1, 0, 0, 1, 1, 1, 1)$.
  - <u>ABSG:</u> Output bit sequence is $(0, 1, 1, 1, 0, 0, 1, 1, 1, 1, 0, 1)$.

Alternatively, BSG and ABSG algorithms can be viewed as two-step algorithms as well. In other words, one can show that BSG (resp. ABSG) algorithm is equivalent to the successive application of two algorithms, namely algorithm $\mathcal{A}$ and algorithm $\mathcal{B}$ (resp. $\mathcal{C}$), whose definitions are given below (see Fig. 1). Although this partitioning is an artificial construction, it provides a deeper insight about BSG's and ABSG's statistical properties and ease of operation.

*Remark 2.1:* We use algorithm $\mathcal{A}$ in order to derive the periodicity properties and output rate results in this paper, so the results we find are valid for both BSG and ABSG algorithms.

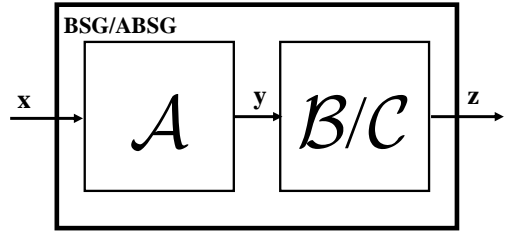

Fig. 1. Block Diagram Representation of BSG and ABSG as two step algorithms.

*Definition 2.1:* Input sequence of algorithm $\mathcal{A}$ is defined as $\mathbf{x} = (x_1, x_2, \ldots)$, where $x_i \in \{0, 1\}$.

*Definition 2.2:* $\mathbf{y} \triangleq \mathcal{A}(\mathbf{x})$, where $\mathbf{y}$ is the internal state of BSG and ABSG algorithms and $y_i \in \{\varnothing, 0, 1\}$, $1 \leq i \leq N$, $i \in \mathbb{Z}^+$. The action of algorithm $\mathcal{A}$ is defined via the mapping $\mathcal{M}$:

$$y_i = \mathcal{M}(y_{i-1}, x_i), \quad 1 \leq i \leq N, \ i \in \mathbb{Z}^+,$$

with the initial condition $y_0 = \varnothing$. The mapping $\mathcal{M}$ is given in Table I [2].

TABLE I
TRANSITION TABLE OF ALGORITHM $\mathcal{A}$

| $y_{i-1} \backslash x_i$ | 0 | 1 |
|---|---|---|
| $\varnothing$ | 0 | 1 |
| 0 | $\varnothing$ | 0 |
| 1 | 1 | $\varnothing$ |

*Remark 2.2:* BSG and ABSG algorithms produce an output bit at time instant $i$ if and only if $y_i = \varnothing$.

*Definition 2.3:* $\mathbf{z} \triangleq \mathcal{B}(\mathbf{y})$, where $\mathbf{z}$ is the output sequence of BSG algorithm; such that, action of algorithm $\mathcal{B}$ is given as follows:

$$z_j = \begin{cases} 0, & \text{if } y_i = \varnothing \text{ and } y_{i-2} = \varnothing, \\ 1, & \text{if } y_i = \varnothing \text{ and } y_{i-2} \neq \varnothing, \end{cases}$$

where $j \leq i$ and $i, j \in \mathbb{Z}^+$.

*Definition 2.4:* $\mathbf{z} \triangleq \mathcal{C}(\mathbf{y})$ where $\mathbf{z}$, is the output sequence of ABSG algorithm; such that, action of algorithm $\mathcal{C}$ is given as follows:

$$z_j = \begin{cases} y_{i-1}, & \text{if } y_i = \varnothing \text{ and } y_{i-2} = \varnothing, \\ \bar{y}_{i-1}, & \text{if } y_i = \varnothing \text{ and } y_{i-2} \neq \varnothing, \end{cases}$$

where $j \leq i$ and $i, j \in \mathbb{Z}^+$.

*Remark 2.3:* One can show that $\mathcal{B}(\mathcal{A}(\mathbf{x}))$ (resp. $\mathcal{C}(\mathcal{A}(\mathbf{x}))$) is equivalent to the BSG (resp. ABSG) algorithm; to see this, we refer the interested reader to [2], [3], where $y_i$ is referred to as the "state" of the algorithm at time $i$.

*Definition 2.5:* Given the input $\mathbf{x}_{i+1}^{i+j}$ and the state $y_i$ at some time $i$ ($i \in \mathbb{N}$), we use $\mathcal{M}^j(\cdot, \cdot)$ to denote the equivalent mapping of applying $\mathcal{M}(\cdot, \cdot)$ $j$ successive times beginning from time $i+1$ to the input $(x_{i+1}, x_{i+2}, \ldots, x_{i+j})$. Hence, we recursively define

$$\mathcal{M}^j\left(y_i, \mathbf{x}_{i+1}^{i+j}\right) = \mathcal{M}\left(\mathcal{M}^{j-1}\left(y_i, \mathbf{x}_{i+1}^{i+j-1}\right), x_{i+j}\right), \quad j \in \mathbb{Z}^+, \tag{1}$$

where $\mathcal{M}^1\left(y_i, \mathbf{x}_{i+1}^{i+1}\right) = \mathcal{M}\left(y_i, x_{i+1}\right) = y_{i+1}$, $i \in \mathbb{N}$.

*Definition 2.6:* We define $\mathcal{S}$ as the set of all possible ordered state values at a specific time:

$$\mathcal{S} \triangleq \left\{ \left(\varnothing, 0, 1\right)^T, \left(\varnothing, 1, 0\right)^T, \left(0, \varnothing, 1\right)^T, \left(0, 1, \varnothing\right)^T, \left(1, \varnothing, 0\right)^T, \left(1, 0, \varnothing\right)^T \right\}.$$

For any $a \in \mathcal{S}$, $a\left(k\right) \in \{\varnothing, 0, 1\}$ denotes the $k$-th element of $a$, $1 \leq k \leq 3$.

*Definition 2.7:* Using time $i \in \mathbb{N}$ as the reference point, given the input $\mathbf{x}_{i+1}^{i+j} \in \{0, 1\}^j$ and the possible ordered state values $\vec{s}_i^{\,i} \in \mathcal{S}$ at time $i$, the vector of ordered states at time $i + j$ is denoted by $\vec{s}_i^{\,i+j} \in \mathcal{S}$, $j \in \mathbb{Z}^+$, where

$$\vec{s}_i^{\,i+j} = \begin{bmatrix} \mathcal{M}^j\left(\vec{s}_i^{\,i}\left(1\right), \mathbf{x}_{i+1}^{i+j}\right) \\ \mathcal{M}^j\left(\vec{s}_i^{\,i}\left(2\right), \mathbf{x}_{i+1}^{i+j}\right) \\ \mathcal{M}^j\left(\vec{s}_i^{\,i}\left(3\right), \mathbf{x}_{i+1}^{i+j}\right) \end{bmatrix}. \tag{2}$$

*Remark 2.4:* We use the convention of $\vec{s}_i = \vec{s}_i^{\,i}$ for all $i \in \mathbb{N}$. Also, note that, for all $i$, $j$, $\vec{s}_i^{\,i+j}$ is a function of both $\mathbf{x}_{i+1}^{i+j}$ and $\vec{s}_i$, but this dependency is not explicitly specified in the notation for the sake of convenience.

*Definition 2.8:* We define the mapping

$$\vec{\mathcal{M}}^j\left(\cdot, \cdot\right) : \mathcal{S} \times \{0, 1\}^j \longmapsto \mathcal{S},$$

such that, for all $i \in \mathbb{N}$, $j \in \mathbb{Z}^+$,

$$\vec{s}_i^{\,i+j} = \vec{\mathcal{M}}^j\left(\vec{s}_i, \mathbf{x}_{i+1}^{i+j}\right),$$

where $\vec{s}_i^{\,i+j}$ is given in (2).

*Remark 2.5:* One way to interpret the mapping $\vec{\mathcal{M}}^j\left(\cdot, \cdot\right)$ is to view it as a permutation on $\vec{s}_i$ as a function of $\mathbf{x}_{i+1}^{i+j}$. In particular, beginning from time $i \geq 0$, given the input $\mathbf{x}_{i+1}^{i+j}$, $\vec{\mathcal{M}}^j$ produces $\vec{s}_i^{\,i+j}$ which is a permuted version of $\vec{s}_i \in \mathcal{S}$. Hence, for fixed input $\mathbf{x}$ (of length-$j$), $\vec{\mathcal{M}}^j$ is a permutation (for all $j$). Next, recall that the set of all permutations on 3 letters forms a group under composition of mappings; in algebra, this is a well-known group, called the "symmetric group of degree 3" and denoted by $S_3$, which is of cardinality $3! = 6$ and known to be non-abelian [11]. We heavily use this interpretation in the proofs of the some of the results presented in the subsequent sections[2].

*Definition 2.9:* Given an input sequence $\mathbf{x}$ (of length-$j$), the *permutation order* of the corresponding mapping $\vec{\mathcal{M}}^j\left(\cdot, \mathbf{x}\right) : \mathcal{S} \mapsto \mathcal{S}$ is the *order* of the corresponding permutation in $S_3$ [11].

Note that, given a length-$j$ input $\mathbf{x}$ and some $\vec{s} \in \mathcal{S}$ on which $\vec{\mathcal{M}}^j$ operates,

- the permutation order of $\vec{\mathcal{M}}^j$ is 1 if it is the identity mapping,
- the permutation order of $\vec{\mathcal{M}}^j$ is 2 if it swaps two elements of $\vec{s}$ and preserves the location of the remaining element,
- the permutation order of $\vec{\mathcal{M}}^j$ is 3 if it changes the locations of all 3 elements of $\vec{s}$.

We have completed defining basic concepts about BSG and ABSG algorithms; in the subsequent sections, we derive some fundamental properties of BSG and ABSG algorithms related to the rate and periodicity.

## III. Deterministic Setup

In this section, we analyze the behavior of BSG and ABSG algorithms for the case of $m$-sequence inputs. We present some basic results in Section III-A. In section III-B, we prove important properties of BSG and ABSG for the case of $m$-sequence inputs, which are used in Section III-C, where upper and lower bounds for the expected periods of BSG and ABSG are derived.

### A. General Properties

In this part, we state some fundamental results which we frequently use throughout the rest of Section III.

*Lemma 3.1:* Given the state $y_i$ at time $i$, $\mathcal{M}\left(y_i, x_{i+1}\right)$ is a one-to-one mapping on $x_{i+1}$ for all $i \in \mathbb{N}$.

*Proof:* Lemma 3.1 is clear from the definition of the mapping $\mathcal{M}\left(\cdot, \cdot\right)$ given in Table I. ∎

*Remark 3.1:* From Lemma 3.1, it is obvious that for every distinct input (resp. output) sequence to algorithm $\mathcal{A}$, there is a unique output (resp. input) sequence.

*Remark 3.2:* Given a length-$j$ input $\mathbf{x}$, the mapping $\vec{\mathcal{M}}^j\left(\cdot, \mathbf{x}\right)$ and the corresponding equivalent permutation $\theta \in S_3$, we have [11]

- the permutation order of $\vec{\mathcal{M}}^j$ is 1 or 3 if and only if $\theta$ is an even permutation,
- the permutation order of $\vec{\mathcal{M}}^j$ is 2 if and only if $\theta$ is an odd permutation.

*Remark 3.3:* [11] Given $\theta, \psi \in S_3$, we have

- if both $\theta$ and $\psi$ are even permutations, then both $\theta \circ \psi$ and $\psi \circ \theta$ are even permutations,

---

[2]In this paper, we employ $S_3$ with a slight abuse in notation: It is customary in algebra to define the symmetric group $S_3$ on pre-defined triplets, whereas in this work, the elements of the group $S_3$ operate on vectors of length-3 which constitute the set $\mathcal{S}$.

- if both $\theta$ and $\psi$ are odd permutations, then both $\theta \circ \psi$ and $\psi \circ \theta$ are even permutations,
- if either "$\theta$ is odd and $\psi$ is even" or "$\theta$ is even and $\psi$ is odd", then both $\theta \circ \psi$ and $\psi \circ \theta$ are odd permutations.

*Lemma 3.2:* Given a length-$N$ input $\mathbf{x}$, we have

1) if $N$ is even, permutation order of $\bar{\mathcal{M}}^N (\cdot, \mathbf{x})$ is 1 or 3,
2) if $N$ is odd, permutation order of $\bar{\mathcal{M}}^N (\cdot, \mathbf{x})$ is 2.

*Proof:* First, we recall the mapping $\mathcal{M}(y, x)$ for any $x \in \{0, 1\}$ from Table I. Clearly, Table I implies that the permutation order of $\bar{\mathcal{M}}^1$ is 2. Hence, from Remark 3.2, we note that the corresponding permutation in $S_3$ is odd. Next, we analyze the mapping $\mathcal{M}^2 (y, \mathbf{x})$ for any length-2 $\mathbf{x} \in \{0, 1\}^2$ in Table II. We conclude that the permutation order of $\bar{\mathcal{M}}^2$ is 1 or 3. Again, from Remark 3.2, we note that the corresponding permutation in $S_3$ is even. Next, we define $\theta$ as the permutation in $S_3$ which corresponds to the mapping $\bar{\mathcal{M}}^N$ for a fixed input. If $N$ is even, then $\theta$ can be expressed as a product of even permutations, which yields an even permutation (Remark 3.3). If $N$ is odd, then $\theta$ can be expressed as a product of multiple even permutations and a single odd permutation, which yields an odd permutation (Remark 3.3). Per Remark 3.2, this, in return, implies that the permutation order of $\bar{\mathcal{M}}^N$ is 1 or 3 (resp. 2) if $N$ is even (resp. odd). ∎

TABLE II
$\mathcal{M}^2(y, \mathbf{x})$ for any length-2 $\mathbf{x}$

| $y \backslash \mathbf{x}$ | 00 | 01 | 10 | 11 |
|---|---|---|---|---|
| $\varnothing$ | $\varnothing$ | 0 | 1 | $\varnothing$ |
| 0 | 0 | 1 | $\varnothing$ | 0 |
| 1 | 1 | $\varnothing$ | 0 | 1 |

Before finishing this section, we state Lemma 3.3 which is heavily used in the proof of Lemma 3.5. Note that, Lemma 3.3 is originally given in [2]; in this paper, we provide the proof using our notation and setup.

*Lemma 3.3:* [2] Given $\mathbf{x}_1^i$ for any $i > 1$, let $\mathbf{x}_{k+1}^i$ be the $k$-shifted version of $\mathbf{x}_1^i$ for some $k \in \{1, 2, \dots, i-1\}$. Then,

$$\left[ \mathcal{M}^i(\varnothing, \mathbf{x}_1^i) = \mathcal{M}^{i-k}(\varnothing, \mathbf{x}_{k+1}^i) \right] \iff \left[ \mathcal{M}^k(\varnothing, \mathbf{x}_1^i) = \varnothing \right]. \tag{3}$$

*Proof:* First, we prove forward statement. We can express $\mathcal{M}^i(\varnothing, \mathbf{x}_1^i)$ with the following form:

$$\mathcal{M}^i(\varnothing, \mathbf{x}_1^i) = \mathcal{M}^{i-k}(\mathcal{M}^k(\varnothing, \mathbf{x}_1^k), \mathbf{x}_{k+1}^i).$$

If $\mathcal{M}^i(\varnothing, \mathbf{x}_1^i) = \mathcal{M}^{i-k}(\varnothing, \mathbf{x}_{k+1}^i)$, Remark 3.1 implies that $\mathcal{M}^k(\varnothing, \mathbf{x}_1^k) = \varnothing$.

Converse statement of (3) is trivial, since

$$\mathcal{M}^i(\varnothing, \mathbf{x}_1^i) = \mathcal{M}^{i-k}(\mathcal{M}^k(\varnothing, \mathbf{x}_1^k), \mathbf{x}_{k+1}^i) = \mathcal{M}^{i-k}(\varnothing, \mathbf{x}_{k+1}^i).$$

∎

Next, we proceed with proving results on periodicity properties.

### B. Periodicity Results for Maximal-Length Sequences

Throughout this section, we assume that $\mathbf{x}$ is generated by a length $L$-LFSR for a *given* primitive feedback polynomial. Hence, $\mathbf{x}$ is a $m$-sequence (maximal length sequence) and one period of it is denoted by $\mathbf{x}_1^T$, where $T = 2^L - 1$ is the period for a $m$-sequence [1].

*Definition 3.1:* Given a length $L$-LFSR, the set of all possible initial states of the form $\mathbf{x}_1^L$, except all zero case, is denoted by $\mathcal{X} = \{0, 1\}^L - \{0\}^L$.

*Remark 3.4:* Obviously, $|\mathcal{X}| = T$.

*Definition 3.2:* $\mathcal{Y} \triangleq \{\mathbf{y} | \mathbf{y} = \mathcal{A}(\mathbf{x})\}$, where $\mathbf{x}$ is generated by a length-$L$ LFSR, for all $\mathbf{x}_1^L \in \mathcal{X}$ as the initial state.

*Remark 3.5:* For every initial state of LFSR, such that $\mathbf{x}_1^L \in \mathcal{X}$, there exists a unique output [1]. Since $|\mathcal{X}| = T$, Remark 3.1 implies that $|\mathcal{Y}| = T$.

*Definition 3.3:*

$$\mathcal{Y}_A \triangleq \{\mathbf{y} \in \mathcal{Y} | y_T = \varnothing\}, \quad \mathcal{Y}_B \triangleq \{\mathbf{y} \in \mathcal{Y} | y_T \neq \varnothing\}.$$

*Remark 3.6:* Using Definition 3.2 and 3.3 we clearly have

$$\mathcal{Y}_A \cap \mathcal{Y}_B = \{ \; \}, \quad \mathcal{Y}_A \cup \mathcal{Y}_B = \mathcal{Y}.$$

In other words, $\mathcal{Y}_A$ and $\mathcal{Y}_B$ are mutually exclusive, collectively exhaustive subsets of $\mathcal{Y}$.

In the following proposition, we state some fundamental properties regarding periods of the elements of $\mathcal{Y}_A$ and $\mathcal{Y}_B$; recall that for algorithm $\mathcal{A}$, we have the initial condition $y_0 = \varnothing$.

*Proposition 3.1:*

**i)** Every $\mathbf{y} \in \mathcal{Y}_A$ is periodic with $T$.

ii) Every $\mathbf{y} \in \mathcal{Y}_B$ is periodic with $2T$.

   *Proof:*

i) First, note that the sequence $\mathbf{y}_1^T$ is generated using the input $\mathbf{x}_1^T$ with the initial condition $y_0 = \varnothing$. Now, for all $\mathbf{y} \in \mathcal{Y}_A$, we have $y_T = \varnothing$. Because $\mathbf{x}$ is periodic with $T$, we have $\mathbf{x}_{T+1}^{2T} = \mathbf{x}_1^T$. Moreover, since $\mathbf{y}_{T+1}^{2T}$ is produced by $\mathbf{x}_{T+1}^{2T}$ with the initial condition $y_T = \varnothing$, we necessarily have $\mathbf{y}_{T+1}^{2T} = \mathbf{y}_1^T$ due to the one-to-one property stated in Remark 3.1. Hence the proof of the first part.

ii) Given a fixed input $\mathbf{x}_1^T$, we know that the permutation order of $\bar{\mathcal{M}}^T \left( \cdot, \mathbf{x}_1^T \right)$ is 2 since $T = 2^L - 1$ is odd (Lemma 3.2). Let $\theta \in S_3$ denote the corresponding permutation to $\bar{\mathcal{M}}^T \left( \cdot, \mathbf{x}_1^T \right)$ for fixed $\mathbf{x}$. Since permutation order of $\bar{\mathcal{M}}^T$ is 2, $\theta$ is odd (Remark 3.2). Next, note that the permutation in $S_3$ corresponding to $\bar{\mathcal{M}}^T \left( \cdot, \mathbf{x}_{T+1}^{2T} \right)$ is also equal to $\theta$ since $\mathbf{x}$ is $T$-periodic. Now, observe that any odd permutation in $S_3$ is the inverse of itself since if a permutation in $S_3$ is odd this means it preserves the location of one element and swaps the locations of the remaining two. This implies $\theta \circ \theta = e$, where $e$ is the identity element in $S_3$. Hence, $\bar{\mathcal{M}}^{2T} \left( \vec{s}, \mathbf{x}_1^{2T} \right) = \vec{s}$ for any $\vec{s} \in \mathcal{S}$ and $T$-periodic $\mathbf{x}$. Therefore, $y_{2T} = \varnothing$ since $y_0 = \varnothing$ per assumption. This, using similar arguments of the proof of part (i), implies that $\mathbf{y} \in \mathcal{Y}_B$ is always $2T$-periodic. ∎

   *Remark 3.7:* As a direct consequence of Remark 3.6 and Proposition 3.1, we observe that given $\mathbf{x}$ as input, algorithm $\mathcal{A}$ generates *exactly* two kinds of output sequences in terms of their periods, namely the ones with period $T$ and $2T$.

   Following lemma is an interesting property of sequences in $\mathcal{Y}_B$, that will be helpful in the proofs Lemma 3.5 and Lemma 3.6.

   *Lemma 3.4:* For all $\mathbf{y} \in \mathcal{Y}_B$ and for all $i \in \{1, 2, \ldots, T\}$, we cannot have $(y_i = \varnothing)$ *and* $(y_{T+i} = \varnothing)$.

   *Proof:* Statement is trivially true if $i = T$, from the definition of $\mathcal{Y}_B$. Furthermore, since we cannot have $y_1 = \varnothing$ from the definition of $\mathcal{M}(\cdot, \cdot)$, statement is true for $i = 1$ as well. Next, we consider $1 < i < T$ and follow proof by contradiction. Suppose, the statement is false for some $\mathbf{y} \in \mathcal{Y}_B$ and some $i \in \{2, 3, \ldots, T-1\}$, i.e., we have $y_i = y_{T+i} = \varnothing$. Using this and the fact that $\mathbf{x}$ is $T$-periodic, we have

$$\varnothing = y_{2T} = \mathcal{M}^{T-i} \left( y_{T+i}, \mathbf{x}_{T+i+1}^{2T} \right) = \mathcal{M}^{T-i} \left( \varnothing, \mathbf{x}_{T+i+1}^{2T} \right) = \mathcal{M}^{T-i} \left( \varnothing, \mathbf{x}_{i+1}^{T} \right) = \mathcal{M}^{T-i} \left( y_i, \mathbf{x}_{i+1}^{T} \right) = y_T,$$

where we also used the one-to-one property of $\mathcal{M}(\cdot, \cdot)$. This yields a contradiction since $\mathbf{y} \in \mathcal{Y}_B$ and $y_T \neq \varnothing$ by definition. ∎

   *Lemma 3.5:* Given an arbitrary $m$-sequence $\mathbf{x}$ such that $\mathbf{y} = \mathcal{A}(\mathbf{x})$,

i) if $\mathbf{y} \in \mathcal{Y}_A$, then

$$[y_k = \varnothing] \implies [\mathcal{A}((\mathbf{x}, k)) \in \mathcal{Y}_A];$$

ii) if $\mathbf{y} \in \mathcal{Y}_B$, then

$$[y_j = \varnothing] \implies [\mathcal{A}((\mathbf{x}, k)) \in \mathcal{Y}_B],$$

   where $k \equiv j \mod T$.

   *Proof:* See Appendix I. ∎

   *Lemma 3.6:*

i) Given an arbitrary $m$-sequence $\mathbf{x}$ such that $\mathbf{y} = \mathcal{A}(\mathbf{x}) \in \mathcal{Y}_A$, and for any given $\tilde{\mathbf{y}} \in \mathcal{Y}_A$, there exists a unique $k \in \{0, 1, 2, \ldots, T-1\}$ such that $\tilde{\mathbf{y}} = \mathcal{A}((\mathbf{x}, k))$ and $y_k = \varnothing$.

ii) Given an arbitrary $m$-sequence $\mathbf{x}$ such that $\mathbf{y} = \mathcal{A}(\mathbf{x}) \in \mathcal{Y}_B$, and for any given $\tilde{\mathbf{y}} \in \mathcal{Y}_B$, there exists a unique $k \in \{0, 1, 2, \ldots, T-1\}$ and a unique $j \in \{0, 1, 2, \ldots, 2T-1\}$ such that $\tilde{\mathbf{y}} = \mathcal{A}((\mathbf{x}, k))$ and $y_j = \varnothing$, where $j \equiv k \mod T$.

   *Proof:* See Appendix II. ∎

   *Theorem 3.1:* Given an arbitrary $m$-sequence $\mathbf{x}$ such that $\mathbf{y} = \mathcal{A}(\mathbf{x})$,

i) if $\mathbf{y} \in \mathcal{Y}_A$, then

$$\mathcal{Y}_A = \{\mathcal{A}((\mathbf{x}, k)) \,|\, y_k = \varnothing, \ 0 \leq k < T\}; \tag{4}$$

ii) if $\mathbf{y} \in \mathcal{Y}_B$, then

$$\mathcal{Y}_B = \{\mathcal{A}((\mathbf{x}, k)) \,|\, y_j = \varnothing, \ 0 \leq k < T, \ j \equiv k \mod T\}. \tag{5}$$

   *Proof:* Part (i) of the theorem is obvious using the first parts of Lemmas 3.5 and 3.6. Likewise, part (ii) of the theorem is obvious using the second parts of Lemmas 3.5 and 3.6. ∎

   *Remark 3.8:* From Theorem 3.1, we note the following results:

i) All $\mathbf{y} \in \mathcal{Y}_A$ are "properly"-shifted versions of each other. To be more precise, all the $\mathbf{y}$'s in $\mathcal{Y}_A$ can be formed by "$\varnothing$-shift"ing an arbitrary $\mathbf{y}_A^{ref} \in \mathcal{Y}_A$.

ii) All $\mathbf{y} \in \mathcal{Y}_B$ are "properly"-shifted versions of each other. To be more precise, all the $\mathbf{y}$'s in $\mathcal{Y}_B$ can be formed by "$\varnothing$-shift"ing an arbitrary $\mathbf{y}_B^{ref} \in \mathcal{Y}_B$.

Here, a "$\varnothing$-shift" of a sequence $\mathbf{y}$ refers to some $(\mathbf{y}, k)$ where $y_k = \varnothing$.

The significance of Remark 3.8 is the following: we can completely characterize the set $\mathcal{Y}_A$ in terms of "$\varnothing$-shifts" of any $\mathbf{y} \in \mathcal{Y}_A$ and similarly $\mathcal{Y}_B$ can be completely characterized in terms of "$\varnothing$-shifts" of any $\mathbf{y} \in \mathcal{Y}_B$. This point turns out to be helpful in deriving results related to cardinalities of $\mathcal{Y}_A$ and $\mathcal{Y}_B$, which will be investigated next[3]. First, we provide the following definition:

*Definition 3.4:*

- Given any $\mathbf{y} \in \mathcal{Y}_A$, $\mathcal{T}_A$ is the number of $\varnothing$'s within one period; i.e., $\mathcal{T}_A \overset{\triangle}{=} \mathcal{W}_\varnothing \left( \mathbf{y}_1^T \right)$ where $\mathbf{y} \in \mathcal{Y}_A$.
- Given any $\mathbf{y} \in \mathcal{Y}_B$, $\mathcal{T}_B$ is the number of $\varnothing$'s within one period; i.e., $\mathcal{T}_B \overset{\triangle}{=} \mathcal{W}_\varnothing \left( \mathbf{y}_1^{2T} \right)$ where $\mathbf{y} \in \mathcal{Y}_B$.

Following corollary is a direct consequence of Theorem 3.1.

*Corollary 3.1:*

$$|\mathcal{Y}_A| = \mathcal{T}_A, \quad |\mathcal{Y}_B| = \mathcal{T}_B.$$

Corollary 3.1 is one of the key results of this paper. Without this result, at first sight, it may look like the cardinalities of $\mathcal{Y}_A$ and $\mathcal{Y}_B$ are about the same [2]; however, as we show in the subsequent sections, this is indeed not the case. In fact, the cardinality of $\mathcal{Y}_A$ (resp. $\mathcal{Y}_B$) is equal to the length of the output sequence $\mathbf{z}$ produced by the ABSG algorithm where $\mathbf{z} = \mathcal{C}\left(\mathbf{y}\right)$, $\mathbf{y} \in \mathcal{Y}_A$ (resp. $\mathbf{y} \in \mathcal{Y}_B$); replacing $\mathcal{C}$ with $\mathcal{B}$, same argument holds for the BSG algorithm as well. Next we proceed with Corollary 3.2, which, together with Corollary 3.1, constitute our main tools in deriving bounds on the periods of the output sequences of BSG and ABSG algorithms.

*Corollary 3.2:*

$$\mathcal{T}_A + \mathcal{T}_B = T. \tag{6}$$

Note that, Corollary 3.2 is a direct consequence of Corollary 3.1 and Remarks 3.5 and 3.6.

From Corollary 3.2, it is obvious that a lower bound for $\mathcal{T}_A$ directly implies an upper bound for $\mathcal{T}_B$ and vice versa. In the next section, we derive lower and upper bounds for $\mathcal{T}_A$ and $\mathcal{T}_B$.

## C. Bounds

Before proceeding any further, we note the following fundamental observation, stated in Remark 3.9, about the period of the output sequence $\mathbf{z}$ for both BSG and ABSG algorithms.

*Remark 3.9:* Combining Proposition 3.1, Definition 3.4, and noting that both BSG and ABSG produce an output bit at time instant $i$ if and only if $y_i = \varnothing$, we conclude that the period of the output sequence $\mathbf{z}$ divides $\mathcal{T}_A$ (resp. $\mathcal{T}_B$) if it is produced by some $\mathbf{y} \in \mathcal{Y}_A$ (resp. $\mathbf{y} \in \mathcal{Y}_B$).

Now, we note our fundamental assumption: We assume that the output sequence $\mathbf{z}$ of ABSG and BSG algorithms has no subperiod (for an experimental justification, see [2], [3]); i.e., the quantity $\mathcal{T}_A$ (resp. $\mathcal{T}_B$) is the *least period* of $\mathbf{z}$ if the corresponding input sequence $\mathbf{y} \in \mathcal{Y}_A$ (resp. $\mathbf{y} \in \mathcal{Y}_B$). Next, we state some auxiliary results that are helpful in deriving bounds on $\mathcal{T}_A$ and $\mathcal{T}_B$.

*Lemma 3.7:* For any three consecutive "run"s[4] in $\mathbf{x}$, we observe at least one $\varnothing$ in the corresponding $\mathbf{y}$ sequence.

*Proof:* Suppose we have runs $r_1, r_2, r_3$ in $\mathbf{x}$ sequence, after the $l^{\text{th}}$ entry:

$$\mathbf{x} = (\dots, b^{r_1}, \bar{b}^{r_2}, b^{r_3}, \dots).$$

Then, we have following alternatives from the definition of algorithm $\mathcal{A}$:

**i)** $y_l = \varnothing \Rightarrow$ if $r_1 \geq 2$, then $y_{l+2} = \varnothing$; if $r_1 = 1$, then $y_{l+r_1+r_2+1} = \varnothing$;

**ii)** $y_l = b \Rightarrow y_{l+1} = \varnothing$;

**iii)** $y_l = \bar{b} \Rightarrow y_{l+r_1+1} = \varnothing$;

where $b \in \{0, 1\}$. ∎

*Remark 3.10:* Given an $(\mathbf{x}, \mathbf{y})$ pair, such that $\mathbf{y} = \mathcal{A}\left(\mathbf{x}\right)$, Lemma 3.7 implies a bound on the minimum number of $\varnothing$'s in $\mathbf{y}$ given the number of runs in $\mathbf{x}$.

*Proposition 3.2:*

$$\lceil \frac{2^L}{6} \rceil \leq \mathcal{T}_A \leq 2^{L-1} - 1. \tag{7}$$

*Proof:* We begin with the lower bound. From the definition of $\mathcal{Y}_A$, we know, by assumption, that $\mathbf{x}$, which forms $\mathbf{y} \in \mathcal{Y}_A$, is a $m$-sequence, for which the total number of runs is $2^{L-1}$ [1]. Partitioning all the runs of $\mathbf{x}$ into groups of 3 and using Lemma 3.7, we obtain $\lfloor \frac{2^{L-1}}{3} \rfloor \leq \mathcal{W}_\varnothing(\mathbf{y}_1^T)$. In addition, $y_T = \varnothing$, since $\mathbf{y} \in \mathcal{Y}_A$. Thus, $\lfloor \frac{2^{L-1}}{3} \rfloor + 1 = \lceil \frac{2^L}{6} \rceil \leq \mathcal{W}_\varnothing(\mathbf{y}_1^T)$ which forms our lower bound. Upper bound directly follows from the definition of algorithm $\mathcal{A}$ since there can be at most 1 bit of output (i.e., one instance of $\varnothing$) per 2 bits of input $\mathbf{x}$ (equivalently 2 bits of $\mathbf{y}$) and the length of $\mathbf{x}$ within one period is $2^L - 1$ [1]. ∎

---

[3]A related discussion on this issue was also provided in [2] without a rigorous proof.

[4]For a precise definition of a "run", see [1].

*Corollary 3.3:*

$$2^{L-1} \leq \mathcal{T}_B \leq 2^L - 1 - \lceil \frac{2^L}{6} \rceil. \tag{8}$$

Corollary 3.3 directly follows from Proposition 3.2 and Corollary 3.2.

*Corollary 3.4:* Assuming that all initial states of the LFSR are equally likely, we have

$$\Pr\left(\mathbf{y} \in \mathcal{Y}_A\right) = \frac{\mathcal{T}_A}{T}, \quad \Pr\left(\mathbf{y} \in \mathcal{Y}_B\right) = \frac{\mathcal{T}_B}{T}$$

Corollary 3.4 is obvious using Corollary 3.1 and Corollary 3.2.

*Remark 3.11:* Letting $\mathcal{T}_{\mathbf{z}}$ denote the average period of $\mathbf{z}$ (where the probability space is induced by all possible equally-likely initial states of LFSR, excluding the state of all zeros), using Corollary 3.4 and Remark 3.9, we have

$$\mathcal{T}_{\mathbf{z}} = \Pr\left(\mathbf{y} \in \mathcal{Y}_A\right) \mathcal{T}_A + \Pr\left(\mathbf{y} \in \mathcal{Y}_B\right) \mathcal{T}_B = \frac{\mathcal{T}_A^2 + \mathcal{T}_B^2}{\mathcal{T}_A + \mathcal{T}_B}. \tag{9}$$

Following bounds on $\mathcal{T}_{\mathbf{z}}$ are direct consequences of Corollary 3.2, Proposition 3.2 and Corollary 3.3.

*Corollary 3.5:*

$$\frac{(2^{L-1}-1)^2 + (2^{L-1})^2}{2^L - 1} \leq \mathcal{T}_{\mathbf{z}} \leq \frac{\lceil \frac{2^L}{6} \rceil^2 + (2^L - 1 - \lceil \frac{2^L}{6} \rceil)^2}{2^L - 1}. \tag{10}$$

The result (10) follows from straightforward calculus, where we perform constrained optimization with the cost function (9), subject to constraints (6,7,8).

*Remark 3.12:* For large values of $L$, (10) may be approximated by the following inequality:

$$\frac{9}{18} 2^L < \mathcal{T}_{\mathbf{z}} < \frac{13}{18} 2^L.$$

*Remark 3.13:* If the input sequence $\mathbf{x}$ is such that $\mathbf{y} = \mathcal{A}\left(\mathbf{x}\right) \in \mathcal{Y}_A$ (resp. $\mathbf{y} = \mathcal{A}\left(\mathbf{x}\right) \in \mathcal{Y}_B$), then the output rates of both BSG and ABSG algorithms (recall the definitions of $\mathcal{B}$ and $\mathcal{C}$) is given by $\frac{\mathcal{T}_A}{T}$ (resp. $\frac{\mathcal{T}_B}{2T}$). Further investigation of the *output rate* is worth pursuing, which constitutes the topic of Section IV.

## IV. PROBABILISTIC SETUP

In this section, we analyze the output rate of ABSG and BSG algorithms under the assumption that the input is a stochastic process (hence probabilistic setup). In particular, we assume that the input sequence is an evenly distributed binary sequence. Since all pseudo random number generators aim to produce sequences that "look" truly random, quantifying the behavior of BSG and ABSG algorithms with evenly distributed input sequences helps us to have a better understanding of these two algorithms.

Since the output rate is directly determined by the number of $\varnothing$'s in the output sequence of algorithm $\mathcal{A}$, denoted by $\{Y_i\}$ (given the length of the input sequence $\{X_i\}$) for both BSG and ABSG algorithms, analyzing the probabilistic behavior of $\{Y_i\}$ suffices to quantify the rate distribution. In order to achieve this task, we initially focus on the distribution of the internal state variables, $\{Y_i\}$, in Section IV-A. In particular, we derive the marginal probability distribution $\Pr\left(Y_i\right)$ and the conditional probability distribution $\Pr\left(Y_i | \mathbf{Y}_0^{i-1}\right)$ for some $i \geq 1$. Using these results for a fixed length input, we calculate the probability mass function of the output length (i.e., given $N$, the number of $\varnothing$'s in $\mathbf{Y}_1^N$) in Section IV-B which directly implies the rate distribution in the probabilistic setup. As a result, we derive the mean and variance of the output rate. In Section IV-C, we extend our analysis to include the asymptotic behavior of the rate; in particular, we show that the output rate is concentrated around its mean with exponential tightness.

### A. Distribution of the Internal State Variables

*Definition 4.1:*

$$\alpha_n \triangleq \Pr(Y_n = \varnothing), \quad \beta_n \triangleq \Pr(Y_n = 0), \quad \theta_n \triangleq \Pr(Y_n = 1).$$

*Theorem 4.1:* For algorithm $\mathcal{A}$, if the input $\{X_i\}$ is an i.i.d Bernoulli process with probability $1/2$, then for $n \in \mathbb{Z}^+$, the following four statements hold

$$\alpha_{2n} = \frac{1}{3} + \frac{2}{3}2^{-2n}, \tag{11}$$

$$\beta_{2n} = \theta_{2n} = \frac{1}{3} - \frac{1}{3}2^{-2n}, \tag{12}$$

$$\alpha_{2n+1} = \frac{1}{3} - \frac{1}{3}2^{-2n}, \tag{13}$$

$$\beta_{2n+1} = \theta_{2n+1} = \frac{1}{3} + \frac{1}{6}2^{-2n}. \tag{14}$$

*Proof:* See the Appendix III. ∎

Next, we concentrate on the conditional probability distribution $\Pr\left(Y_i | \mathbf{Y}_0^{i-1}\right)$. Using Definition 2.2, we immediately observe the following result:

*Corollary 4.1:* From Table I, we see that if the input sequence $\{X_i\}$ is evenly distributed, the internal state sequence $\{Y_i\}$ is a Markov process with memory one and the initial condition $Y_0 = \varnothing$, which implies

$$\Pr\left(Y_i | \mathbf{Y}_0^{i-1}\right) = \Pr\left(Y_i | Y_{i-1}\right),$$

where, for all $i > 0$,

$$\Pr\left(Y_i = \varnothing | Y_{i-1} \neq \varnothing\right) = \frac{1}{2}, \qquad \Pr\left(Y_i \neq \varnothing | Y_{i-1} \neq \varnothing\right) = \frac{1}{2},$$
$$\Pr\left(Y_i = \varnothing | Y_{i-1} = \varnothing\right) = 0, \qquad \Pr\left(Y_i \neq \varnothing | Y_{i-1} = \varnothing\right) = 1.$$

### B. Distribution of the Output Length and the Rate

Since we aim to characterize the distribution of the number of $\varnothing$'s in $\mathbf{Y}_1^N$ (given $N$), we first define an auxiliary random sequence $\{Q_i\}$ for the sake of convenience.

*Definition 4.2:* At each time instant $i \geq 0$, the random variable $Q_i$ is defined as

$$Q_i = \begin{cases} 1, & \text{if } Y_i = \varnothing, \\ 0, & \text{otherwise.} \end{cases}$$

*Remark 4.1:* Using Corollary 4.1 and Definition 4.2, we observe that if the input sequence $\{X_i\}$ is evenly distributed, the sequence $\{Q_i\}$ is a Markov process with memory one and the initial condition $Q_0 = 1$, which implies

$$\Pr\left(Q_i | \mathbf{Q}_0^{i-1}\right) = \Pr\left(Q_i | Q_{i-1}\right),$$

where, for all $i > 0$,

$$\Pr\left(Q_i = 1 | Q_{i-1} = 0\right) = \frac{1}{2}, \qquad \Pr\left(Q_i = 0 | Q_{i-1} = 0\right) = \frac{1}{2},$$
$$\Pr\left(Q_i = 1 | Q_{i-1} = 1\right) = 0, \qquad \Pr\left(Q_i = 0 | Q_{i-1} = 1\right) = 1.$$

Next, we derive the probability mass function (pmf) of $\mathcal{W}_\varnothing\left(\mathbf{Y}_1^N\right)$ which will yield the probabilistic behavior of the output rate of BSG and ABSG algorithms. For the sake of convenience, we use the following definition.

*Definition 4.3:* $H \triangleq \mathcal{W}_\varnothing(\mathbf{Y}_1^N) = \mathcal{W}_1(\mathbf{Q}_1^N)$.

*Theorem 4.2:* The probability mass function of $H$ is given by

$$\Pr(H = k) = \begin{cases} 2^{-N+1}, & \text{for } k = 0, \\[2mm] \binom{N-k-1}{k}2^{-(N-k-1)} + \binom{N-k-1}{k-1}2^{-(N-k)}, & \text{for } 0 < k < \frac{N}{2} \text{ and } k \in \mathbb{Z}^+, \\[2mm] 2^{-\frac{N}{2}}, & \text{for } N \text{ even and } k = \frac{N}{2}. \end{cases} \tag{15}$$

*Proof:* See Appendix IV ∎

Next, we derive the mean and variance of $H$.

*Proposition 4.1:*

$$\mathrm{E}[H] = \frac{N}{3} - \frac{2}{9} + \frac{2}{9}\left(-\frac{1}{2}\right)^N. \tag{16}$$

*Proof:* We have

$$
\begin{aligned}
\mathrm{E}\left[H\right] &= \sum_{i=1}^{N}\Pr\left(Q_i=1\right)=\sum_{i=1}^{N}\alpha_i \\
&= \sum_{i=1}^{N}\left[\frac{1}{3}+\frac{2}{3}\left(-\frac{1}{2}\right)^i\right] \\
&= \frac{N}{3}+\frac{2}{3}\left[\sum_{i=0}^{N}\left(-\frac{1}{2}\right)^i-1\right]=\frac{N}{3}-\frac{2}{3}+\frac{2}{3}\frac{1-\left(-\frac{1}{2}\right)^{N+1}}{\frac{3}{2}} \\
&= \frac{N}{3}-\frac{2}{9}+\frac{2}{9}\left(-\frac{1}{2}\right)^N,
\end{aligned}
\tag{17}
$$

where (17) follows from (11) and (13). ∎

*Remark 4.2:* The output rate of an algorithm is the reciprocal of the number of input bits needed to produce one output bit, so the output rate of BSG and ABSG algorithms is $H/N$, whose expected value is given by

$$
\mathrm{E}[H/N]=\frac{\mathrm{E}[H]}{N}=\frac{1}{3}-\frac{2}{9N}+\frac{2}{9N}(-\frac{1}{2})^N.
\tag{18}
$$

Note that, (18) gives the analytical expression for the expected value of the rate, with the asymptotic value of $1/3$; the asymptotic result has also been provided in [2], [3].

*Remark 4.3:* We note that, under the evenly-distributed input *approximation* in the deterministic setup, (18) implies that $\mathcal{T}_A\approx\frac{T}{3}$ which is also justified by experiments (recall the results of Section III-B). Since $\mathcal{T}_A+\mathcal{T}_B=T$ (Corollary 3.2), (18) also implies $\mathcal{T}_B\approx\frac{2T}{3}$. Moreover, recalling Remark 3.13, this observation implies that the rate in the deterministic setup is about $\frac{1}{3}$ under the evenly-distributed input approximation. Also, from (9), we note that $T_{\mathbf{z}}\approx\frac{5T}{9}$ under this approximation, which implies that the lower bound of Remark 3.12 is tighter than its upper bound counterpart.

*Proposition 4.2:*

$$
\mathrm{Var}\left(H\right)=\sigma_H^2=\frac{2N}{27}+\frac{2}{81}+\left(\frac{4N}{27}+\frac{2}{81}\right)\left(-\frac{1}{2}\right)^N-\frac{4}{81}\left(\frac{1}{4}\right)^N.
\tag{19}
$$

*Proof:* See the Appendix V. ∎

Next, we aim to find out the concentration of the rate around its mean. Since the actual distribution is difficult to handle, we analyze it asymptotically, which is the topic of the next section.

### C. Asymptotic Behavior and Bounds

In the discussions and developments presented in this section, we heavily make use of "attributes", "recurrent events" and their properties. A comprehensive reading about this subject is given in Chapter XIII of [12].

We term $\varepsilon$ to be an "attribute" of the finite sequence $(A_{i_1}, A_{i_2}, \ldots, A_{i_n})$ if it is uniquely determined whether this sequence has, or has not the characteristic $\varepsilon$. Then, the statement "$\varepsilon$ occurs at the $n$-th place in the sequence $\{A_{i_j}\}$" is equivalent to saying "the subsequence $(A_{i_1}, A_{i_2}, \ldots, A_{i_n})$ has the attribute $\varepsilon$" [12].

*Definition 4.4:* [12] The attribute $\varepsilon$ defines a *recurrent event* if:

- In order that $\varepsilon$ occurs at the $n$-th and $(n+m)$-th place of the sequence $\left\{A_{i_j}\right\}_{j=1}^{n+m}$ it is necessary and sufficient that $\varepsilon$ occurs at the last place in each of the two subsequences $\left\{A_{i_j}\right\}_{j=1}^{n}$ and $\left\{A_{i_j}\right\}_{j=n+1}^{n+m}$.
- Whenever this is the case, we have

$$
\Pr\left(A_{i_1}, A_{i_2}, \ldots, A_{i_{n+m}}\right)=\Pr\left(A_{i_1}, A_{i_2}, \ldots, A_{i_n}\right)\Pr\left(A_{i_{n+1}}, A_{i_{n+2}}, \ldots, A_{i_{n+m}}\right).
$$

*Definition 4.5:* We define the attribute $\zeta$ such that it is said to occur at the $n$-th place in the (potentially infinite) sequence $\{Q_i\}$ if $Q_n=1$.

*Lemma 4.1:* The attribute $\zeta$ defines a recurrent event.

*Proof:* First, we note that, in order to have $\zeta$ occurring at the $n$-th and $(n+m)$-th places of the sequence $(Q_1, Q_2, \ldots, Q_{n+m})$, it is necessary and sufficient to have $Q_n=Q_{n+m}=1$, which implies that $\zeta$ occurs at the last places of the two subsequences $(Q_1, Q_2, \ldots, Q_n)$ and $(Q_{n+1}, Q_{n+2}, \ldots, Q_{n+m})$. Furthermore, because of the Markovian property of $\{Q_i\}$ and the initial condition $Q_0=1$, we have

$$
\begin{aligned}
&\Pr\left[\zeta \text{ occurs at the $n$-th and $(n+m)$-th places of }(Q_1,Q_2,\ldots,Q_{n+m})\right] \\
&=\Pr\left(Q_{n+m}=1,Q_n=1|Q_0=1\right)=\Pr\left(Q_{n+m}=1|Q_n=1\right)\cdot\Pr\left(Q_n=1|Q_0=1\right) \\
&=\Pr\left[\zeta \text{ occurs at the $n$-th place of }(Q_1,Q_2,\ldots,Q_n)\right] \\
&\quad\cdot\Pr\left[\zeta \text{ occurs at the $(n+m)$-th place of }(Q_{n+1},Qn+2,\ldots,Q_{n+m})\right],
\end{aligned}
$$

which implies that $\zeta$ is recurrent. ∎

*Lemma 4.2:* The recurrent event $\zeta$ is *persistent*.

*Proof:* We recall that [12] $\zeta$ is persistent if $\sum_{n=1}^{\infty} f_n = 1$, where

$$f_n \triangleq \Pr\left(\zeta \text{ occurs for the first time at the } n\text{-th trial}\right). \tag{20}$$

Hence,

$$f_n = \Pr\left(Q_1 = 0, Q_2 = 0, \ldots, Q_{n-1} = 0, Q_n = 1 | Q_0 = 1\right).$$

Thus, clearly

$$f_1 = 0. \tag{21}$$

For $n > 1$, using the Markovian property of $\{Q_i\}$ and Corollary 4.1 we obtain

$$
\begin{aligned}
f_n &= \underbrace{\Pr\left(Q_1 = 0 | Q_0 = 1\right)}_{1} \left[\prod_{i=2}^{n-1} \underbrace{\Pr\left(Q_i = 0 | Q_{i-1} = 0\right)}_{1/2}\right] \underbrace{\Pr\left(Q_n = 1 | Q_{n-1} = 0\right)}_{1/2} \\
&= 2^{-(n-1)},
\end{aligned} \tag{22}
$$

which implies

$$\sum_{n=1}^{\infty} f_n = \sum_{n=2}^{\infty} \left(\frac{1}{2}\right)^{n-1} = \frac{1}{2} \sum_{n=0}^{\infty} \left(\frac{1}{2}\right)^{n} = \frac{1/2}{1 - 1/2} = 1.$$

∎

*Theorem 4.3:* Asymptotically, as $N \to \infty$, $H$ is Gaussian distributed with mean $N/3$ and variance $2N/27$.

*Proof:* First, we note that $H$ represents the number of occurrences of $\zeta$ in the first $N$ trials. Next, we introduce the random variable $T$ such that

$$\Pr\left(T = n\right) = f_n,$$

where $f_n$ is defined in (20) and its value is given in (21,22). Note that, $T$ can also be referred to as the *recurrence time* of $\zeta$. Let $\mu_T$ and $\sigma_T^2$ represent the mean and variance of $T$, respectively. We know that, if $\mu_T, \sigma_T^2 < \infty$, as $N \to \infty$, $H \sim \mathcal{N}\left(\frac{N}{\mu_T}, \frac{N\sigma_T^2}{\mu_T^3}\right)$ ([12], p. 297, Theorem 1). Before proceeding further, recall the following standard results from Calculus: Given $\alpha < 1$, we have

$$\sum_{i=0}^{\infty} \alpha^i = \frac{1}{1-\alpha}, \quad \sum_{i=0}^{\infty} i\alpha^i = \frac{\alpha}{(1-\alpha)^2}, \quad \sum_{i=0}^{\infty} i^2\alpha^i = \frac{\alpha\left(1+\alpha\right)}{(1-\alpha)^3}. \tag{23}$$

Using (21,22,23), we get

$$\mu_T = \mathbb{E}\left(T\right) = \sum_{i=1}^{\infty} if_i = \sum_{i=2}^{\infty} \frac{i}{2^{i-1}} = 2\sum_{i=0}^{\infty} i\left(\frac{1}{2}\right)^i - 1 = 2\frac{1/2}{\left(1-1/2\right)^2} - 1 = 3,$$

$$\sigma_T^2 = \mathbb{E}\left(T^2\right) - \mu_T^2 = \sum_{i=1}^{\infty} i^2 f_i - \mu_T^2 = \sum_{i=2}^{\infty} i^2 \frac{1}{2^{i-1}} - 9 = 2\sum_{i=0}^{\infty} i^2 \left(\frac{1}{2}\right)^i - 10 = 2\frac{(1/2)(3/2)}{\left(1-1/2\right)^3} - 10 = 2,$$

which are obviously finite, consequently $\frac{N}{\mu_T} = \frac{N}{3}$ and $\frac{N\sigma_T^2}{\mu_T^3} = \frac{2N}{27}$. ∎

*Corollary 4.2:* We asymptotically have

$$\Pr\left(|H - \mathbb{E}\left[H\right]| > \gamma \mathbb{E}\left[H\right]\right) \approx 2\mathbb{Q}\left(\sqrt{\frac{3N}{2}}\gamma\right) < \frac{2}{\sqrt{2\pi}\gamma(3N/2)^{1/2}} e^{-\frac{3N}{4}\gamma^2},$$

where $\mathbb{Q}\left(x\right) \triangleq \int_x^{\infty} \frac{1}{\sqrt{2\pi}} e^{-t^2/2} \, dt$. Hence, $H$ is asymptotically exponentially tight around its mean value $N/3$.

Corollary 4.2 directly follows from Theorem 4.3, the definition of the Q-function and the well-known upper bound of $\mathbb{Q}\left(\alpha\right) < \frac{1}{\sqrt{2\pi}\alpha} e^{-\alpha^2/2}$.

Next, we illustrate the aforementioned results via an experimental study. In Fig. 2, we compare the actual distribution of $H = \mathcal{W}_\varnothing\left(\mathbf{Y}_1^N\right) = \mathcal{W}_1\left(\mathbf{Q}_1^N\right)$ (given in Theorem 4.2) with the asymptotically-converging Gaussian distribution (given in Theorem 4.3). In the left panel, we compare the corresponding c.d.f.s (cumulative distribution functions); in the right panel, the relative entropy (Kullback-Leibler) distance is used as the basis of comparison[5]. We note that even for remarkably small values of $N$ for cryptographic purposes (e.g., $N = 100$), the asymptotic Gaussian approximation is valid in practice. Recall that the case of length-$N$ roughly corresponds to a length-$\log N$ LFSR in practical implementations; thus, our experiments reveal that convergence to Gaussian approximation is remarkably fast.

---

[5] Recall that for two distributions $p(t)$ and $q(t)$, the relative entropy between $p$ and $q$ is given by $D\left(p||q\right) = \int_t p\left(t\right) \log \frac{p(t)}{q(t)} \, dt$.

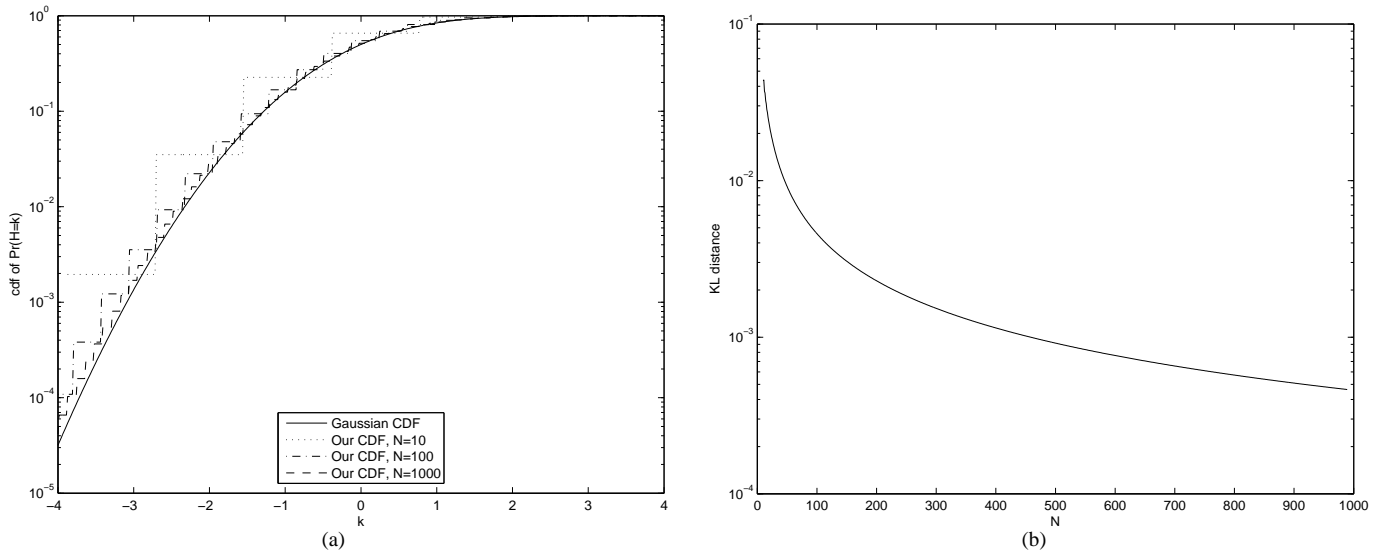

Fig. 2. Comparison of the actual distribution of $H = \mathcal{W}_{\varnothing}\left(\mathbf{Y}_1^N\right) = \mathcal{W}_1\left(\mathbf{Q}_1^N\right)$ given in Theorem 4.2 and the asymptotically-valid Gaussian distribution given in Theorem 4.3; (a) the comparison is done via plotting c.d.f.s (cumulative distribution functions); the dotted, dash-dotted, and dashed lines show the actual c.d.f. of $H$ for $N = 10$, $N = 100$, $N = 1000$, respectively; the solid line represents the Gaussian approximation; (b) we compare the actual distribution of $H$ and the Gaussian distribution in the sense of relative entropy (also known as Kullback-Leibler distance) as a function of the length of the sequence, $N$.

## V. Conclusion

In this paper, we develop a further theoretic understanding of BSG and ABSG algorithms and analytically quantify periodicity and output rate properties. As far as the input sequence is concerned, we consider both deterministic and probabilistic setups; all of our results hold both for BSG and ABSG algorithms. In the deterministic case, we derive fundamental results on periodicity properties, where we assume that the input is a $m$-sequence. We prove that there are *exactly* two different disjoint sets of output sequences; in addition, any element in one of these two sets is a proper shift of any other element in the same set. Moreover, by using this partitioning, we derive bounds on the expected output period under the no subperiod assumption. In the probabilistic setup, we assume that the input is a realization of an i.i.d Bernoulli process with probability $1/2$. We derive the probability mass function of the number of output bits given the input length and analytically derive the output rate. Moreover, we prove that the aforementioned distribution converges to a Gaussian distribution as the sample size tends to infinity. Further, we use this result to show that the output rate is exponentially-concentrated around $1/3$, which is a notable property of BSG and ABSG.

## Acknowledgments

The authors are grateful to Yağmur Denizhan and Ali Aydın Selçuk for various discussions and suggestions which helped to improve this paper.

## Appendix I
### Proof of Lemma 3.5

**i)** Since Lemma 3.3 is valid for all $i$, $k$, $k < i$, it directly implies that

$$[y_k = \varnothing] \quad \Longleftrightarrow \quad [\tilde{\mathbf{y}} = (\mathbf{y}, k)],$$

where $\tilde{\mathbf{y}} \overset{\triangle}{=} \mathcal{A}\left((\mathbf{x}, k)\right)$. Hence, in order to prove part (i) of Lemma 3.5, we need to show that

$$[\tilde{\mathbf{y}} = (\mathbf{y}, k)] \quad \Longrightarrow \quad [\tilde{\mathbf{y}} \in \mathcal{Y}_A].$$

Note that, $\tilde{\mathbf{y}}$ is generated by $(\mathbf{x}, k)$, which is a $m$-sequence as well [1]. This means

$$[\tilde{y}_T = \varnothing] \quad \Longrightarrow \quad [\tilde{\mathbf{y}} \in \mathcal{Y}_A].$$

Thus, in order to prove part (i) of Lemma 3.5, it is sufficient to show that

$$[\tilde{\mathbf{y}} = (\mathbf{y}, k)] \quad \Longrightarrow \quad [\tilde{y}_T = \varnothing]. \tag{I-1}$$

Now, we have

$$\varnothing \quad = \quad y_k, \tag{I-2}$$

$$= \quad y_{T+k}, \tag{I-3}$$

$$= \quad \tilde{y}_T, \tag{I-4}$$

where (I-2) follows from the assumption of part (i) of Lemma 3.5, (I-3) follows from the fact that $\mathbf{y} \in \mathcal{Y}_A$ and is $T$-periodic, (I-4) follows from the definition of $\tilde{\mathbf{y}}$. Hence, (I-1) follows, that completes proof of part (i) of Lemma 3.5.

**ii)** Using arguments similar to those of part (i) and using the fact that $\mathbf{x}$ is $T$-periodic, one can show that in order to prove part (ii) of Lemma 3.5, it is sufficient to show that

$$[\tilde{\mathbf{y}} = (\mathbf{y}, j)] \quad \Longrightarrow \quad [\tilde{y}_T \neq \varnothing], \tag{I-5}$$

where $k \equiv j \mod T$. Note that, since $\mathbf{y} \in \mathcal{Y}_B$, we assume without loss of generality that $0 \le j < 2T$ since $\mathbf{y}$ is $2T$-periodic per definition. In other words, we prove the claim for the first period and in that case it trivially holds for all the other periods. Thus, we only need to deal with two cases: $j = k$ and $j = k + T$, where $0 \le k < T$. First, assume $y_k = \varnothing$; then, we have

$$\varnothing \quad = \quad y_k, \tag{I-6}$$

$$= \quad y_{2T+k}, \tag{I-7}$$

$$= \quad \tilde{y}_{2T}, \tag{I-8}$$

where (I-6) follows from the assumption of $y_k = \varnothing$, (I-7) follows from the fact that $\mathbf{y} \in \mathcal{Y}_B$ and is $2T$-periodic, (I-8) follows from the definition of $\tilde{\mathbf{y}}$. Per Lemma 3.4, (I-8) implies $\tilde{y}_T \neq \varnothing$. Hence, the statement (I-5) is valid for $j = k$, $0 \le k < T$. Using similar arguments with $j = k + T$ instead of $j = k$, it is obvious that (I-5) is also valid for the case of $j = T + k$, $0 \le k < T$ (where $y_{T+k} = \varnothing$). $\qquad \square$

# APPENDIX II
## PROOF OF LEMMA 3.6

We begin with defining the concepts that will be used in the proofs of both part (i) and part (ii).

First, we define the permutations $e, \alpha, \beta, \eta, \theta, \gamma \in S_3$ , acting on the triplet $(x_1, x_2, x_3)$ [11],

$$
\begin{array}{llllll}
e: & x_1 \mapsto x_1 & \eta: & x_1 \mapsto x_1 & \theta: & x_1 \mapsto x_2 & \beta: & x_1 \mapsto x_2 & \alpha: & x_1 \mapsto x_3 & \gamma: & x_1 \mapsto x_3 \\
& x_2 \mapsto x_2 & & x_2 \mapsto x_3 & & x_2 \mapsto x_1 & & x_2 \mapsto x_3 & & x_2 \mapsto x_1 & & x_2 \mapsto x_2 \\
& x_3 \mapsto x_3 & & x_3 \mapsto x_2 & & x_3 \mapsto x_3 & & x_3 \mapsto x_1 & & x_3 \mapsto x_2 & & x_3 \mapsto x_1
\end{array}
$$

Next, we concentrate on the first $T$ samples of $\mathbf{x}$ and $\tilde{\mathbf{x}} \overset{\triangle}{=} (\mathbf{x}, k)$. Here, since both $\mathbf{x}$ and $\tilde{\mathbf{x}}$ are $m$-sequences (i.e., both are $T$-periodic), we have $\mathbf{x}_1^T = (\mathbf{x}_1^k, \mathbf{x}_{k+1}^T)$ and $\tilde{\mathbf{x}}_1^T = (\tilde{\mathbf{x}}_1^{T-k}, \tilde{\mathbf{x}}_{T-k+1}^T) = (\mathbf{x}_{k+1}^T, \mathbf{x}_1^k)$. Fixing $\mathbf{x}_1^k$ and $\mathbf{x}_{k+1}^T$, we define the mappings

$$\varphi \; : \; \mathcal{S} \mapsto \mathcal{S}, \quad \text{and} \quad \psi \; : \; \mathcal{S} \mapsto \mathcal{S},$$

such that

$$[\vec{u} = \varphi(\vec{s})] \iff \left[ \vec{u} = \vec{\mathcal{M}}^k(\vec{s}, \mathbf{x}_1^k) \right],$$

$$[\vec{u} = \psi(\vec{s})] \iff \left[ \vec{u} = \vec{\mathcal{M}}^{T-k}(\vec{s}, \mathbf{x}_{k+1}^T) \right],$$

for any $\vec{s}, \vec{u} \in \mathcal{S}$ (recall Definition 2.6). Next, recalling Remark 2.5, we note that $\varphi, \psi \in S_3$.

Now, we proceed with the proof Lemma 3.6. First, note that for the case of $\tilde{\mathbf{y}} = \mathbf{y}$, the lemma trivially holds with $k = 0$ recalling the assumption of $y_0 = \varnothing$ for $\mathcal{A}$. Next, we consider the case of $\tilde{\mathbf{y}} \neq \mathbf{y}$ and discuss each part of the lemma separately.

**i)** Since $\tilde{\mathbf{y}} \in \mathcal{Y}_A$ per assumption of the part (i) of lemma, it should be produced by a $m$-sequence that is a shifted version of $\mathbf{x}$, where $\mathbf{y} = \mathcal{A}(\mathbf{x})$. Since any $m$-sequence for a given feedback polynomial is a shifted version of another $m$-sequence for the same feedback polynomial [1], there exists some $k \in \{1, 2, \dots, T-1\}$ such that $\tilde{\mathbf{y}} = \mathcal{A}((\mathbf{x}, k))$; furthermore, such a $k$ is unique because $\mathcal{A}(\cdot)$ is one-to-one (recall Remark 3.1). Hence, proving part (i) of Lemma 3.6 reduces to showing $y_k = \varnothing$.

Next, we define the auxiliary state variable $\vec{s}_0 \overset{\triangle}{=} (\varnothing, 0, 1)^T \in \mathcal{S}$ and use it as the reference point. Note that since $\mathbf{y}, \tilde{\mathbf{y}} \in \mathcal{Y}_A$, we have $\varnothing = y_T = \tilde{y}_T$. Also, since $T = 2^L - 1$ is odd, permutation order of $\vec{\mathcal{M}}^T$ is 2 (Lemma 3.2). This implies that

$$\vec{\mathcal{M}}^T(\vec{s}_0, \mathbf{x}_1^T) = \vec{\mathcal{M}}^T(\vec{s}_0, \tilde{\mathbf{x}}_1^T) = (\varnothing, 1, 0)^T. \tag{II-1}$$

Moreover, for fixed $\mathbf{x}_1^T$ (hence for fixed $\tilde{\mathbf{x}}_1^T$ since $\tilde{\mathbf{x}}_1^T = \left(\mathbf{x}_{k+1}^T, \mathbf{x}_1^k\right)$), (II-1) is equivalent to

$$\psi \circ \varphi \left(\vec{s}_0\right) = \varphi \circ \psi \left(\vec{s}_0\right) = \left(\varnothing, 1, 0\right)^T.$$

Due to the definition of $\eta \in S_3$, this further implies

$$\psi \circ \varphi = \varphi \circ \psi = \eta.$$

Since $\eta \neq e$, $\varphi$ and $\psi$ cannot be inverses of each other. Furthermore, $S_3$ is non-abelian, so we necessarily need to have one of the two following cases:

- Case 1: $\varphi = e$ and $\psi = \eta$ ,
- Case 2: $\psi = e$ and $\varphi = \eta$ .

Now, observe that in both cases, $\varphi$ preserves the location of the first element, i.e., the first element of $\varphi \left(\vec{s}_0\right)$ is equal to $\varnothing$. Hence, we necessarily have $y_k = \varnothing$, which completes the proof of part (i).

**ii)** Since $\tilde{\mathbf{y}} \in \mathcal{Y}_B$ per assumption of the part (ii) of lemma, it should be produced by a $m$-sequence that is a shifted version of $\mathbf{x}$, where $\mathbf{y} = \mathcal{A}\left(\mathbf{x}\right)$. This implies that there exists some $k \in \{1, 2, \ldots, T-1\}$ such that $\tilde{\mathbf{y}} = \mathcal{A}\left((\mathbf{x}, k)\right)$; furthermore, such a $k$ is unique because $\mathcal{A}\left(\cdot\right)$ is one-to-one (recall Remark 3.1). Also, recall that Lemma 3.4 implies both $y_k$ and $y_{T+k}$ cannot be $\varnothing$ at the same time. Hence, the remaining task is to prove $y_j = \varnothing$ for $j \equiv k \mod T$ for some *unique* $j \in \{1, 2, \ldots, 2T-1\}$ since $\mathcal{Y}_B$ is $2T$-periodic (i.e., it is sufficient to show either $y_k = \varnothing$ or $y_{T+k} = \varnothing$).

Now, using the same reference point $\vec{s}_0$ as in part (i), and noting that since $\mathbf{y}, \tilde{\mathbf{y}} \in \mathcal{Y}_B$, we have $y_T \neq \varnothing$ and $\tilde{y}_T \neq \varnothing$. Also, since $T$ is odd, permutation order of $\vec{\mathcal{M}}^T$ is 2 (Lemma 3.2), which implies for $\mathbf{y}$

$$\vec{\mathcal{M}}^T \left(\vec{s}_0, \mathbf{x}_1^T\right) = (0, \varnothing, 1)^T \quad \text{or} \quad \vec{\mathcal{M}}^T \left(\vec{s}_0, \mathbf{x}_1^T\right) = (1, 0, \varnothing)^T, \tag{II-2}$$

and for $\tilde{\mathbf{y}}$

$$\vec{\mathcal{M}}^T \left(\vec{s}_0, \tilde{\mathbf{x}}_1^T\right) = (0, \varnothing, 1)^T \quad \text{or} \quad \vec{\mathcal{M}}^T \left(\vec{s}_0, \tilde{\mathbf{x}}_1^T\right) = (1, 0, \varnothing)^T. \tag{II-3}$$

Moreover, for fixed $\mathbf{x}_1^T$ (hence for fixed $\tilde{\mathbf{x}}_1^T$ since $\tilde{\mathbf{x}}_1^T = \left(\mathbf{x}_{k+1}^T, \mathbf{x}_1^k\right)$), (II-2) and (II-3) are equivalent to

$$\psi \circ \varphi \left(\vec{s}_0\right) = (0, \varnothing, 1)^T \quad \text{or} \quad \psi \circ \varphi \left(\vec{s}_0\right) = (1, 0, \varnothing)^T,$$

and

$$\varphi \circ \psi \left(\vec{s}_0\right) = (0, \varnothing, 1)^T \quad \text{or} \quad \varphi \circ \psi \left(\vec{s}_0\right) = (1, 0, \varnothing)^T,$$

respectively.

Due to the definition of $\theta \in S_3$ and $\gamma \in S_3$, this can also be rewritten as

$$\psi \circ \varphi = \theta \quad \text{or} \quad \psi \circ \varphi = \gamma.$$

and

$$\varphi \circ \psi = \theta \quad \text{or} \quad \varphi \circ \psi = \gamma.$$

respectively.

Now, the tedious part of the proof begins. We have following four possibilities:

- Case 1, $\varphi \circ \psi = \psi \circ \varphi = \theta$ : Using similar arguments to those used for the proof of part (i), one can show that we either have $(\varphi = e, \psi = \theta)$, or, $(\varphi = \theta, \psi = e)$. If $(\varphi = e, \psi = \theta)$ (resp. $(\varphi = \theta, \psi = e)$), then $y_k = \varnothing$ (resp. $y_{T+k} = \varnothing$).
- Case 2, $\varphi \circ \psi = \psi \circ \varphi = \gamma$ : Using similar arguments to those used for the proof of part (i), one can show that we either have $(\varphi = e, \psi = \gamma)$, or, $(\varphi = \gamma, \psi = e)$. If $(\varphi = e, \psi = \gamma)$ (resp. $(\varphi = \gamma, \psi = e)$), then $y_k = \varnothing$ (resp. $y_{T+k} = \varnothing$).
- Case 3 , $(\psi \circ \varphi = \gamma, \varphi \circ \psi = \theta)$ : Obviously, we have $\varphi \neq e$ and $\psi \neq e$ (suppose not; this means the one which is not equal to $e$ should be equal to both $\gamma$ and $\theta$, which leads to contradiction). Also, noting that

$$[\psi = e] \quad \Longleftrightarrow \quad [\varphi = e],$$

we conclude $\varphi \neq \gamma$; from symmetry, this also means $\psi \neq \gamma$. Similarly, we also see that $\varphi \neq \theta$ and $\psi \neq \theta$. Thus, $\varphi, \psi \notin \{e, \gamma, \theta\}$, i.e., $\varphi, \psi \in \{\eta, \alpha, \beta\}$. Before proceeding, note that the permutation order of $\eta$ (resp. $\alpha$ and $\beta$) is 2 (resp. 3). Now, we have the following alternatives for the "parity" of $k$:

1) $\underline{k \text{ is odd:}}$ The permutation order of $\varphi$ is 2, which directly implies $\varphi = \eta$. Hence, $y_k = \varnothing$.
2) $\underline{k \text{ is even:}}$ In this case, $T - k = 2^L - 1 - k$ is odd, which means the permutation order of $\psi$ is 2, i.e., $\psi = \eta$. Also, note that the permutation order of $\psi \circ \varphi$ is 2 since $T$ is odd, which implies $e = \psi \circ \varphi \circ \psi \circ \varphi$ (see the proof of Proposition 3.1). Thus,

$$\varphi \circ \psi \circ \varphi = \psi^{-1} = \eta^{-1} = \eta.$$

This implies, $y_{T+k} = \varnothing$.

- Case 4 , $(\psi \circ \varphi = \theta, \varphi \circ \psi = \gamma)$ : Using symmetry, the proof of case 3 also applies here.

Hence, we necessarily have either $y_k = \varnothing$ or $y_{T+k} = \varnothing$. $\qquad\square$

## Appendix III
## Proof of Theorem 4.1

In order to prove Theorem 4.1, we first provide Lemmas III.1, III.2 and III.3. We show that Lemma III.1 (resp. Lemma III.2) implies Lemma III.2 (resp. Lemma III.3). Finally, we use Lemma III.3 in the proof of Theorem 4.1. Throughout this section, uppercase boldface letters denote matrices (in contrast with the rest of the paper).

*Lemma III.1:* Given the $k$x$k$ matrix $\mathbf{U}$, where

$$\mathbf{U} \triangleq \begin{bmatrix} \alpha & \alpha & \dots & \alpha \\ \vdots & \vdots & \ddots & \vdots \\ \alpha & \alpha & \dots & \alpha \end{bmatrix}, \tag{III-1}$$

for some $\alpha \in \mathbb{R}$, we have

$$\mathbf{U}^n = (k\alpha)^{n-1}\mathbf{U}, \quad \text{for} \quad n \in \mathbb{Z}^+.$$

*Proof:* We follow proof by induction.

**i)** for $n = 2$:

Note that $\mathbf{U} = \alpha \mathbf{v}\mathbf{v}^T$, where

$$\mathbf{v} \triangleq (1, 1, \dots, 1)^T, \quad \mathbf{v} \in \mathbb{R}^k, \tag{III-2}$$

is a $k$x$1$ vector. As a direct consequence, we can write

$$\mathbf{U}^2 = \alpha^2 \mathbf{v} \underbrace{\mathbf{v}^T \mathbf{v}}_{k} \mathbf{v}^T = k\alpha^2 \mathbf{v}\mathbf{v}^T = k\alpha \underbrace{\alpha \mathbf{v}\mathbf{v}^T}_{\mathbf{U}} = k\alpha\mathbf{U}. \tag{III-3}$$

**ii)** for $n > 2$:

Suppose the claim holds for $n-1$; i.e., $\mathbf{U}^{n-1} = (k\alpha)^{n-2}\mathbf{U}$. Then,

$$\mathbf{U}^n = \mathbf{U}\mathbf{U}^{n-1} = \mathbf{U}(k\alpha)^{n-2}\mathbf{U} = (k\alpha)^{n-2}\mathbf{U}^2. \tag{III-4}$$

Using (III-3) in (III-4), we have

$$\mathbf{U}^n = (k\alpha)^{n-2}(k\alpha)\mathbf{U} = (k\alpha)^{n-1}\mathbf{U}. \tag{III-5}$$

$\blacksquare$

*Lemma III.2:* Given the $k$x$k$ matrix $\mathbf{V}$, where

$$\mathbf{V} \triangleq \begin{bmatrix} \alpha+1 & \alpha & \dots & \alpha & \alpha \\ \alpha & \alpha+1 & \dots & \alpha & \alpha \\ \vdots & \vdots & \ddots & \vdots & \vdots \\ \alpha & \alpha & \dots & \alpha+1 & \alpha \\ \alpha & \alpha & \dots & \alpha & \alpha+1 \end{bmatrix}, \tag{III-6}$$

for some $\alpha \in \mathbb{R}$, we have

$$\mathbf{V}^n = \mathbf{I} + \frac{(k\alpha+1)^n - 1}{k}\mathbf{W},$$

where $\mathbf{W} \triangleq \mathbf{v}\mathbf{v}^T$, $\mathbf{v}$ is defined in (III-2), and $\mathbf{I}$ is the $k \times k$ identity matrix.

*Proof:* Note that $\mathbf{V} = \mathbf{I} + \mathbf{U}$, where

$$\mathbf{U} = \alpha\mathbf{W} = \alpha\mathbf{v}\mathbf{v}^T, \tag{III-7}$$

as defined in (III-1). Hence,

$$
\begin{aligned}
\mathbf{V}^n &= (\mathbf{I}+\mathbf{U})^n = \sum_{i=0}^n \binom{n}{i}\mathbf{U}^i\mathbf{I}^{n-i} = \mathbf{I} + \sum_{i=1}^n \binom{n}{i}\mathbf{U}^i\mathbf{I}^{n-i} = \mathbf{I} + \sum_{i=1}^n \binom{n}{i}\mathbf{U}^i \\
&= \mathbf{I} + \sum_{i=1}^n \binom{n}{i}(k\alpha)^{i-1}\mathbf{U} \\
&= \mathbf{I} + \mathbf{U}\frac{1}{k\alpha}\sum_{i=1}^n \binom{n}{i}(k\alpha)^i = \mathbf{I} + \mathbf{U}\frac{1}{k\alpha}\left[\sum_{i=0}^n \binom{n}{i}(k\alpha)^i - 1\right] = \mathbf{I} + \mathbf{U}\frac{1}{k\alpha}\left[(1+k\alpha)^n - 1\right] \\
&= \mathbf{I} + \frac{(k\alpha+1)^n - 1}{k}\mathbf{W},
\end{aligned}
\tag{III-8, III-9}
$$

where (III-8) follows from Lemma III.1, and (III-9) follows from (III-7). ∎

*Lemma III.3:* Defining

$$\mathbf{A} \overset{\triangle}{=} \begin{bmatrix} 0 & 1 & 1 \\ 1 & 1 & 0 \\ 1 & 0 & 1 \end{bmatrix},$$ (III-10)

for all $n \in \mathbb{Z}^+$ we have

$$\mathbf{A}^{2n} = \mathbf{I}_3 + \frac{1}{3}(2^{2n} - 1)\mathbf{W}_3,$$ (III-11)

$$\mathbf{A}^{2n+1} = \mathbf{A} + \frac{2}{3}(2^{2n} - 1)\mathbf{W}_3,$$ (III-12)

where

$$\mathbf{W}_3 \overset{\triangle}{=} \begin{bmatrix} 1 & 1 & 1 \\ 1 & 1 & 1 \\ 1 & 1 & 1 \end{bmatrix} \quad \text{and} \quad \mathbf{I}_3 \overset{\triangle}{=} \begin{bmatrix} 1 & 0 & 0 \\ 0 & 1 & 0 \\ 0 & 0 & 1 \end{bmatrix}.$$

*Proof:* A straightforward calculation shows

$$\mathbf{A}^2 = \begin{bmatrix} 2 & 1 & 1 \\ 1 & 2 & 1 \\ 1 & 1 & 2 \end{bmatrix},$$

which satisfies (III-11). Hence, $\mathbf{A}^2$ is of the form (III-6) with $k = 3$ and $\alpha = 1$, which also implies that $\mathbf{A}^2 = \mathbf{I}_3 + \mathbf{W}_3$. Now, using Lemma III.2, we have

$$\mathbf{A}^{2n} = \mathbf{I}_3 + \frac{4^n - 1}{3}\mathbf{W}_3 = \mathbf{I}_3 + \frac{1}{3}(2^{2n} - 1)\mathbf{W}_3,$$

which proves (III-11). Next, we note that

$$\mathbf{A}\mathbf{W}_3 = \mathbf{W}_3\mathbf{A} = \begin{bmatrix} 0 & 1 & 1 \\ 1 & 1 & 0 \\ 1 & 0 & 1 \end{bmatrix}\begin{bmatrix} 1 & 1 & 1 \\ 1 & 1 & 1 \\ 1 & 1 & 1 \end{bmatrix} = 2\mathbf{W}_3.$$ (III-13)

Thus, we have

$$\mathbf{A}^{2n+1} = \mathbf{A}^{2n}\mathbf{A} = (\mathbf{I}_3 + \frac{1}{3}(2^{2n} - 1)\mathbf{W}_3)\mathbf{A}$$ (III-14)

$$= \mathbf{A} + \frac{1}{3}(2^{2n} - 1)\mathbf{W}_3\mathbf{A}$$

$$= \mathbf{A} + \frac{2}{3}(2^{2n} - 1)\mathbf{W}_3,$$ (III-15)

where (III-14) and (III-15) follow from (III-11) and (III-13), respectively. Hence, the proof of (III-12). ∎

Next, we proceed with the proof of the theorem. Because $\{X_i\}$ is evenly distributed, we have

$$\Pr(X_n = 0) = \Pr(X_n = 1) = \frac{1}{2}.$$

Using the definition of algorithm $\mathcal{A}$ (see Table I), we write

$$\alpha_n = \frac{1}{2}(\beta_{n-1} + \theta_{n-1}),$$

$$\beta_n = \frac{1}{2}(\alpha_{n-1} + \beta_{n-1}),$$

$$\theta_n = \frac{1}{2}(\alpha_{n-1} + \theta_{n-1}),$$

which implies

$$\mathbf{p}_n = \frac{1}{2}\mathbf{A}\mathbf{p}_{n-1},$$

where $\mathbf{A}$ is defined in (III-10) and $\mathbf{p}_n \overset{\triangle}{=} [\alpha_n, \beta_n, \theta_n]^T$ with the initial condition $\mathbf{p}_0 = [1, 0, 0]^T$. Therefore,

$$\forall n \in \mathbb{Z}^+, \quad \mathbf{p}_n = 2^{-n}\mathbf{A}^n\mathbf{p}_0.$$ (III-16)

Next,

$$\mathbf{p}_{2n} = 2^{-2n}\mathbf{A}^{2n}\mathbf{p}_0 \tag{III-17}$$

$$= 2^{-2n}\left[\mathbf{I}_3 + \frac{1}{3}(2^{2n}-1)\mathbf{W}_3\right]\mathbf{p}_0 \tag{III-18}$$

$$= 2^{-2n}\mathbf{p}_0 + \frac{1}{3}\left(1-2^{-2n}\right)\begin{bmatrix}1\\1\\1\end{bmatrix} = \frac{1}{3}\begin{bmatrix}1+2^{-2n+1}\\1-2^{-2n}\\1-2^{-2n}\end{bmatrix}, \tag{III-19}$$

where (III-17) follows from (III-16), (III-18) follows from (III-11). Hence the proofs of (11) and (12). Similarly,

$$\mathbf{p}_{2n+1} = 2^{-(2n+1)}\mathbf{A}^{2n+1}\mathbf{p}_0 \tag{III-20}$$

$$= 2^{-(2n+1)}\left[\mathbf{A} + \frac{2}{3}\left(2^{2n}-1\right)\mathbf{W}_3\right]\mathbf{p}_0 \tag{III-21}$$

$$= 2^{-(2n+1)}\mathbf{A}\mathbf{p}_0 + \frac{1}{3}\left(1-2^{-2n}\right)\mathbf{W}_3\mathbf{p}_0 = 2^{-(2n+1)}\begin{bmatrix}0\\1\\1\end{bmatrix} + \frac{1}{3}\left(1-2^{-2n}\right)\begin{bmatrix}1\\1\\1\end{bmatrix}$$

$$= \frac{1}{3}\begin{bmatrix}1\\1\\1\end{bmatrix} + 2^{-2n}\left(\begin{bmatrix}0\\1/2\\1/2\end{bmatrix} - \frac{1}{3}\begin{bmatrix}1\\1\\1\end{bmatrix}\right)$$

$$= \frac{1}{3}\begin{bmatrix}1-2^{-2n}\\1+\frac{1}{2}2^{-2n}\\1+\frac{1}{2}2^{-2n}\end{bmatrix}$$

where (III-20) follows from (III-16), (III-21) follows from (III-12). Hence the proofs of (13) and (14). □

## Appendix IV
## Proof of Theorem 4.2

Note that, given any $l \geq 1$, we have

$$\Pr\left(Q_{n+l}=1, Q_{n+l-1}=0, \ldots, Q_{n+1}=0 | Q_n=1\right) = \underbrace{\Pr\left(Q_{n+l}=1 | Q_{n+l-1}=0\right)}_{\frac{1}{2}} \times \left[\prod_{k=n+1}^{n+l-2}\underbrace{\Pr\left(Q_{k+1}=0 | Q_k=0\right)}_{\frac{1}{2}}\right]$$
$$\times \underbrace{\Pr\left(Q_{n+1}=0 | Q_n=1\right)}_{1} \tag{IV-1}$$

$$= 2^{-(l-1)}, \tag{IV-2}$$

where (IV-1) follows from the fact that $\{Q_i\}$ is a Markov process with memory 1. Since

$$\Pr\left(Q_{i+1}=1 | Q_i=0\right) = \Pr\left(Q_{i+1}=0 | Q_i=0\right) = 1/2,$$

we also have

$$\Pr\left(Q_{n+l}=0, Q_{n+l-1}=0, \ldots, Q_{n+1}=0 | Q_n=1\right) = 2^{-(l-1)}. \tag{IV-3}$$

First, considering the trivial case of no "1"s in $\mathbf{Q}_1^N$, applying (IV-3) we get

$$\Pr\left(H=0\right) = 2^{-(N-1)},$$

which constitutes the first line of (15). Next, assuming $\mathcal{W}_1\left(\mathbf{Q}_1^N\right) > 0$, we consider the two following cases (under the assumption that $Q_0 = 1$):

- Case 1 ($Q_N = 1$): Suppose $\mathbf{Q}_1^N \in \{0,1\}^N$ is a sequence with $k$ "1"s, where $Q_N = 1$. This means that we have $k$ "run"s of "0"s between these "1"s; let $l_i - 1$ denote the length of the $i$-th run of "0"s, $1 \leq i \leq k$:

$$\underbrace{1}_{Q_0}, \underbrace{0,0,\ldots,0}_{\text{length } l_1-1}, 1, \underbrace{0,0,\ldots,0}_{\text{length } l_2-1}, 1, 0, \ldots, 0, 1, \underbrace{0,0,\ldots,0}_{\text{length } l_k-1}, \underbrace{1}_{Q_N}$$

Here, note that $\sum_{i=1}^{k} l_i = N$. Then, using this result, the Markovian property of $\{Q_i\}$ and (IV-2) we have

$$\Pr\left(\mathbf{Q}_1^N\right) = \prod_{i=1}^{k} 2^{-(l_i-1)} = 2^{-\left[\left(\sum_{i=1}^{k} l_i\right)-k\right]} = 2^{-(N-k)}.$$

The remaining task in this case is to "count" the number of such $\left\{ \mathbf{Q}_1^N \right\}$ (i.e., the ones with $Q_N = 1$). Per assumption and the description of mapping $\mathcal{M}$, we have $Q_1 = 0$ and $Q_N = 1$, which leaves $N - 2$ symbols. Since we necessarily have a "0" coming after a "1", this means we aim to find the number of different ways to put $k - 1$ patterns of "10" in a sequence of length $N - 2$ if $H = k$. In this case, we have a total of $\binom{N-k-1}{k-1}$ such possibilities. Hence,

$$\Pr\left( H = k \,,\, Q_N = 1 \right) = \binom{N-k-1}{k-1} 2^{-(N-k)} \quad \text{for } 1 \leq k \leq \tfrac{N}{2},\, k \in \mathbb{Z}^+. \tag{IV-4}$$

- Case 2 ($Q_N = 0$): Suppose $\mathbf{Q}_1^N \in \{0,1\}^N$ is a sequence with $k$ "1"s, where $Q_N = 0$. In this case, we have $k + 1$ runs of "0"s. Again, let $l_i$ denote the length of the $i$-th "0" run, $1 \leq i \leq k+1$:

$$\underbrace{1}_{Q_0},\, \underbrace{0,0,\ldots,0}_{\text{length } l_1 - 1},\, 1,\, \underbrace{0,0,\ldots,0}_{\text{length } l_2 - 1},\, 1,0,\ldots,0,1,\, \underbrace{0,0,\ldots,0}_{\text{length } l_k - 1},\, 1,\, \overbrace{\underbrace{0,0,\ldots,0}_{\text{length } l_{k+1} - 1}}^{Q_N}$$

Note that, in contrast with case 1, here $\sum_{i=1}^{k+1} l_i = N + 1$. Using this result, the Markovian property of $\{Q_i\}$, (IV-2), and (IV-3) we have

$$\Pr\left( \mathbf{Q}_1^N \right) = \left[ \prod_{i=1}^{k} 2^{-(l_i - 1)} \right] 2^{-(l_{k+1} - 2)} = 2^{-\left[ \left( \sum_{i=1}^{k+1} l_i \right) - (k+2) \right]} = 2^{-(N-k-1)}.$$

Next, we "count" the number of $\mathbf{Q}_1^N$ such that $Q_N = 0$. Following similar arguments to those of case 1, here we aim to find the number of different ways to put $k$ patterns of "10" in a sequence of length $N - 1$ if $H = k$. In this case, we have a total of $\binom{N-k-1}{k}$ such possibilities. Hence,

$$\Pr\left( H = k \,,\, Q_N = 0 \right) = \binom{N-k-1}{k} 2^{-(N-k-1)} \quad \text{for } 1 \leq k < \tfrac{N}{2},\, k \in \mathbb{Z}^+. \tag{IV-5}$$

Using (IV-4) and (IV-5) in

$$\Pr\left( H = k \right) = \Pr\left( H = k, Q_N = 0 \right) + \Pr\left( H = k, Q_N = 1 \right)$$

we obtain the second line of (15).

Assuming that $N$ is even, the case of $\mathcal{W}_1\left( \mathbf{Q}_1^N \right) = N/2$ deserves separate attention. In this case, observe that we necessarily have $Q_N = 1$, which implies that (IV-5) does not hold. Using $k = N/2$ in (IV-4) yields the third line of (15). $\qquad\square$

## Appendix V
### Proof of Proposition 4.2

First, note that (recalling $Q_0 = 1$ with probability 1 per assumption)

$$\Pr\left( Q_i = 1 \right) = \Pr\left( Q_i = 1 | Q_0 = 1 \right) = \frac{1}{3} + \frac{2}{3}\left( -\frac{1}{2} \right)^i, \quad i \geq 0, \tag{V-1}$$

from (11) and (13). Next, noting that $\Pr\left( Q_j | Q_i \right)$ depends only on $j - i$ for $j \geq i$ from the definition of $\mathcal{M}$, we also have

$$\Pr\left( Q_j = 1 | Q_i = 1 \right) = \Pr\left( Q_{j-i} = 1 | Q_0 = 1 \right) = \frac{1}{3} + \frac{2}{3}\left( -\frac{1}{2} \right)^{j-i}, \quad \text{for } j \geq i \geq 0. \tag{V-2}$$

Using (V-1) and (V-2), we get

$$\begin{aligned}
\Pr\left( Q_i = 1, Q_j = 1 \right) &= \Pr\left( Q_i = 1 | Q_j = 1 \right) \Pr\left( Q_j = 1 \right) = \left[ \frac{1}{3} + \frac{2}{3}\left( -\frac{1}{2} \right)^{j-i} \right] \left[ \frac{1}{3} + \frac{2}{3}\left( -\frac{1}{2} \right)^i \right] \\
&= \frac{1}{9} + \frac{2}{9}\left( -\frac{1}{2} \right)^i + \frac{4}{9}\left( -\frac{1}{2} \right)^j + \frac{2}{9}\left( -\frac{1}{2} \right)^{j-i}, \quad j \geq i \geq 0.
\end{aligned} \tag{V-3}$$

Next, we have

$$
\begin{aligned}
\mathrm{E}\left[H^2\right] &= \mathrm{E}\left[\sum_{i=1}^{N} Q_i\right]^2 = \mathrm{E}\left[\sum_{i=1}^{N} Q_i^2 + \sum_{\forall i,j,i\neq j} Q_j Q_i\right] \\
&= \sum_{i=1}^{N} \mathrm{E}\left[Q_i^2\right] + \sum_{\forall i,j,i\neq j} \mathrm{E}\left[Q_j Q_i\right] = \sum_{i=1}^{N} \mathrm{Pr}\left(Q_i=1\right) + \sum_{\forall i,j,i\neq j} \mathrm{Pr}\left(Q_j=1, Q_i=1\right) \\
&= \sum_{i=1}^{N} \mathrm{Pr}\left(Q_i=1\right) + 2\sum_{i=1}^{N-1}\sum_{j=i+1}^{N} \mathrm{Pr}\left(Q_j=1, Q_i=1\right) \\
&= \sum_{i=1}^{N}\left[\frac{1}{3}+\frac{2}{3}\left(-\frac{1}{2}\right)^i\right] + 2\sum_{i=1}^{N-1}\sum_{j=i+1}^{N}\left[\frac{1}{9}+\frac{2}{9}\left(-\frac{1}{2}\right)^i + \frac{4}{9}\left(-\frac{1}{2}\right)^j + \frac{2}{9}\left(-\frac{1}{2}\right)^{j-i}\right] & \text{(V-4)} \\
&= \frac{N}{3}+\frac{N^2-N}{9}+\frac{2}{3}\sum_{i=1}^{N}\left(-\frac{1}{2}\right)^i + \frac{4}{9}\sum_{i=1}^{N-1}\sum_{j=i+1}^{N}\left[\left(-\frac{1}{2}\right)^i + 2\left(-\frac{1}{2}\right)^j + \left(-\frac{1}{2}\right)^{j-i}\right] \\
&= \frac{N^2+2N}{9}+\frac{2}{9}\left[-1+\left(-\frac{1}{2}\right)^N\right] + \frac{4}{9}\sum_{i=1}^{N-1}\left[(N-i)\left(-\frac{1}{2}\right)^i\right] + \frac{4}{9}\sum_{i=1}^{N-1}\left[2+\left(-\frac{1}{2}\right)^{-i}\right]\left[\sum_{j=i+1}^{N}\left(-\frac{1}{2}\right)^j\right] \\
&= \frac{N^2+2N}{9}+\frac{2}{9}\left[-1+\left(-\frac{1}{2}\right)^N\right] + \frac{4}{9}\sum_{i=1}^{N-1}\left[(N-i)\left(-\frac{1}{2}\right)^i\right] \\
&\quad +\frac{4}{27}\sum_{i=1}^{N-1}\left[2\left(-\frac{1}{2}\right)^N - 2\left(-\frac{1}{2}\right)^i + \left(-\frac{1}{2}\right)^{N-i} - 1\right] \\
&= \frac{N^2+2N}{9}+\frac{2}{9}\left[-1+\left(-\frac{1}{2}\right)^N\right] + \frac{12N-8}{27}\sum_{i=1}^{N-1}\left(-\frac{1}{2}\right)^i - \frac{4}{9}\sum_{i=1}^{N-1}\left[i\left(-\frac{1}{2}\right)^i\right] \\
&\quad +\frac{8(N-1)}{27}\left(-\frac{1}{2}\right)^N + \frac{4}{27}\sum_{i=1}^{N-1}\left(-\frac{1}{2}\right)^{N-i} - \frac{4(N-1)}{27} \\
&= \frac{3N^2+2N-2}{27}+\left(-\frac{1}{2}\right)^N\frac{8N-2}{27}+\frac{12N-4}{27}\sum_{i=1}^{N-1}\left(-\frac{1}{2}\right)^i - \frac{4}{9}\sum_{i=1}^{N-1}\left[i\left(-\frac{1}{2}\right)^i\right] \\
&= \frac{3N^2+2N-2}{27}+\left(-\frac{1}{2}\right)^N\frac{8N-2}{27}+\frac{12N-4}{27}\left[-\frac{1}{3}-\frac{2}{3}\left(-\frac{1}{2}\right)^N\right] - \frac{4}{9}\left[-\frac{2}{9}-\frac{6N-2}{9}\left(-\frac{1}{2}\right)^N\right] \\
&= \frac{9N^2-6N+6}{81}+\left(-\frac{1}{2}\right)^N\frac{8N-2}{27}, & \text{(V-5)}
\end{aligned}
$$

where (V-4) follows from (V-1) and (V-3). Furthermore,

$$
\begin{aligned}
\left(\mathrm{E}\left[H\right]\right)^2 &= \left[\frac{3N-2}{9}+\frac{2}{9}\left(-\frac{1}{2}\right)^N\right]^2 & \text{(V-6)} \\
&= \frac{9N^2-12N+4}{81}+\frac{4}{81}\left(\frac{1}{4}\right)^N+\frac{12N-8}{81}\left(-\frac{1}{2}\right)^N, & \text{(V-7)}
\end{aligned}
$$

where (V-6) follows from (16). Combining (V-5) and (V-7), (19) follows. $\square$